\pdfoutput=1
\documentclass[journal]{IEEEtran}

\usepackage{cite}

\usepackage{amsmath,amssymb}
\usepackage{amsthm}
\usepackage{mathtools}

\usepackage{enumitem}
\usepackage{graphicx}
\graphicspath{{figs/}}
\usepackage{booktabs}
\usepackage{multirow}
\usepackage{xcolor}

\usepackage{stfloats}

\usepackage[hidelinks]{hyperref}

\theoremstyle{plain}
\newtheorem{theorem}{Theorem}
\newtheorem{lemma}[theorem]{Lemma}

\theoremstyle{definition}
\newtheorem{definition}[theorem]{Definition}
\newtheorem{assumption}[theorem]{Assumption}
\theoremstyle{remark}
\newtheorem{remark}[theorem]{Remark}

\makeatletter
\def\thm@space@setup{%
  \thm@preskip=0.9\baselineskip
  \thm@postskip=0.7\baselineskip}
\makeatother

\begin{document}

\title{Recall-Bounded Staleness for Filtered Multi-Tenant ANN Graph
Indexes under Concurrent Updates}

\author{Namrata~Das, Kailas~Devadkar, Anand~Godbole\\
Sardar Patel Institute of Technology, Mumbai, India}

\markboth{Das \MakeLowercase{\textit{et al.}}: Recall-Bounded Staleness for Filtered Multi-Tenant ANN Graph Indexes}%
{Das \MakeLowercase{\textit{et al.}}: Recall-Bounded Staleness for Filtered Multi-Tenant ANN Graph Indexes}

\maketitle

\begin{abstract}
Production vector search must continuously ingest, delete, and re-tenant vectors while serving
low-latency, high-recall, metadata-filtered approximate nearest-neighbor (ANN) queries for many tenants
with per-tenant service-level objectives. Existing graph-based ANN systems address these requirements largely
in isolation, leaving recall during concurrent re-optimization without an analytical bound. We develop a graph-based
index that combines streaming updates, filtered search, and multi-tenancy, with one theoretical result and two
system mechanisms. Our theory decomposes per-tenant recall loss into staleness from updates not yet integrated
into the graph and filtered navigability determined by the local density of points satisfying a tenant's predicate.
The staleness component exhibits anti-amortization on a merge-based graph, integrating a larger batch of inserts
increases recall degradation super-linearly, so larger merges do not reduce total staleness loss. Holding the insert
horizon fixed while increasing merge size $20\times$, we observe a monotonic decrease in terminal recall@10 from
$99.44\%$ to $99.01\%$. For filtered search, the theory gives a parameter-free frontier-mass relationship in local
predicate density and a query-averaged per-tenant recall bound. We instantiate the bound with a filter and
tenant-aware staging cache and a select-not-resize scheduler with per-tenant staleness ceilings. On a FreshDiskANN
substrate, the ceiling eliminates tight-tier SLA violations under the evaluated churn workload, while merge
reordering changes terminal recall by only $0.02$ percentage points. The results show how recall bounds can guide
background maintenance without relying on merge resizing as a recall control mechanism.
\end{abstract}


\IEEEpeerreviewmaketitle

\section{Introduction}
\label{sec:intro}

\IEEEPARstart{A}{widely-held} intuition about maintaining a graph-based approximate-nearest-neighbor (ANN) index under a stream of
inserts is that batching amortizes cost, since folding more pending points into each merge lowers per-point
maintenance overhead. This holds for \emph{throughput}; for \emph{recall} it does not. Integrating a
batch of $c$ inserts into a merge-based graph index degrades disk-graph recall@$k$ \emph{super-linearly} in $c$, so
the total recall loss accumulated over an insertion horizon does not fall with the merge size. Larger, less
frequent merges are no better for recall than smaller, more frequent ones, and on our point estimates they are
worse. We call this \emph{anti-amortization}. In practice only the weak form matters, since re-sizing merges cannot
engineer degradation away and can only \emph{re-prioritize} it, and this motivates the system we build. The design
rests on the \emph{direction} of the effect ($L^*$ non-decreasing in $c$), which we measure on two independent
axes; the magnitude of the super-linearity is less tightly pinned, and its per-merge-to-horizon assembly is a
conservative step we find loose by about $2\times$ (\S\ref{sec:eval-accum}).

Anti-amortization matters because the workloads that most need graph ANN are exactly the ones that mutate
continuously. Production vector search must ingest, delete, and re-tenant vectors while serving low-latency,
high-recall queries that carry metadata predicates, for many tenants each with its own service-level objective
(SLO). Recent work names the pain directly. CANDOR-Bench~\cite{candorbench} observes that under concurrency
``throughput in concurrent settings is often constrained by insertion overhead rather than query latency, which
highlights a mismatch between streaming workloads and designs originally tuned for offline construction.'' The
VDBMS Survey~\cite{vdbmssurvey} places incremental $k$-NN among its open problems: e-commerce and recommender
platforms retrieve an effectively very large $k$ in small increments so that results appear seamlessly, and while
techniques exist for that access pattern,
``it is unclear how to support this search inside vector indexes.'' A VLDB tutorial on filtered vector
search~\cite{filteredsearchtutorial} makes the corresponding point on the predicate axis, giving as a key objective
``the importance of achieving stable recall, ideally in a declarative manner, ensuring consistent recall across
queries.'' And, from industry, Pinecone~\cite{pineconefilter} reports that
``data is continuously inserted and deleted, and metadata may be updated independently of the vector index, yet
filters must be applied accurately and efficiently at query time.''

The literature is heavily worked along each axis \emph{separately} but not at their intersection. Filtered-static
graph ANN (Filtered-DiskANN, ACORN, RNSG, iRangeGraph, STABLE) builds one index over the joint
feature-and-attribute space and does not stream. Streaming graph ANN (FreshDiskANN, SPFresh, Mycelium-Index, and
the recent single-tenant SOTA PipeANN/OdinANN) operates on unfiltered, single-tenant data and provides at best
\emph{empirical} recall stability under churn. Multi-tenant vector search (Curator, Pinecone, GaussDB-Vector) is
sparse; Curator is static and in-memory, Pinecone sits on an IVF substrate with no analytical bound, and none
schedules background work by per-tenant SLA. No prior system occupies their intersection, and none bounds the
recall a tenant loses during live re-optimization of a filtered graph index.

Our contribution occupies the axes those systems leave open, combining streaming, filtered, and multi-tenant
operation with per-tenant SLA scheduling, an analytical bound, and disk-resident scale. On the axes they do target,
PipeANN and OdinANN lead single-tenant streaming and Curator leads static per-tenant recall, and we position
against rather than displace them.

We make one theoretical and two engineering contributions. Our theorem (C1, \S\ref{sec:theorem}) establishes the
anti-amortization bound and generalizes it to the filtered, multi-tenant, deletes/migration setting. Its filtered
term is a \emph{parameter-free} frontier-mass law in the \emph{local} predicate density $\sigma_{\mathrm{loc}}$ that
predicts the recovery scale $L_{95}=1.25\,k/\sigma_{\mathrm{loc}}$ and yields a per-tenant admission test
$1.25\,k/\sigma_{\mathrm{loc}}\le L_{\max}(\ell_\tau)$; evaluated as a query average, it is a conservative
per-tenant recall lower bound. The staleness half is measured at horizon scale rather than extrapolated, with
terminal recall decreasing monotonically as the merge size grows across a $20\times$ range of $c$.

Two engineering designs instantiate the bound. A filter and tenant-aware staging cache (C2, \S\ref{sec:system})
absorbs in-flight inserts, deletes, and tenancy changes on the query path and collapses to the baseline on a single
partition. An SLA-aware, select-not-resize scheduler (C3) honors a per-tenant max-staleness ceiling, re-prioritizing
which pending work each fixed-size merge drains rather than re-sizing the merge that anti-amortization forbids.

On a FreshDiskANN substrate we measure that anti-amortization holds end to end (at a fixed $60$K-insert horizon,
terminal recall@10 falls $99.44\to99.01$ as merges grow $1\text{K}\to20\text{K}$), that re-ordering merges leaves
terminal recall unchanged ($|\Delta\text{recall@10}|=0.02$ pp), that the per-tenant staleness ceiling eliminates
tight-tier SLA violations under churn where a ceiling-less baseline fails, and that the design remains
memory-bounded and recall-correct at $100$K.

\section{System Model and Notation}
\label{sec:model}

\subsection*{Vectors, queries, predicates}
Let $\mathcal{X}\subseteq\mathbb{R}^d$ be the ambient space with metric $\rho$. A point $x\in\mathcal{X}$ carries
metadata $m(x)\in\mathcal{M}$; a \emph{predicate} $p:\mathcal{M}\to\{0,1\}$ selects the filtered subset
$\mathcal{X}_p=\{x:p(m(x))=1\}$, of \emph{global selectivity} $\sigma(p)=|\mathcal{X}_p|/|\mathcal{X}|$. A
\emph{query} is $q=(v,k,p)$ with query vector $v$, neighbor count $k$, and predicate $p$; its ground truth is the
$k$ closest points of $\mathcal{X}_p$ to $v$ under $\rho$. Because filtered-graph recall is governed by predicate
density \emph{near the query} rather than globally (\S\ref{sec:theorem}), we also define the query-joint
\emph{local density} $\sigma_{\mathrm{loc}}(p,v)=\Pr_{x\sim\mathrm{nbhd}(v)}[p(m(x))=1]$, the fraction of $v$'s
metric neighborhood that satisfies $p$.

\subsection*{Tenants and SLOs}
The point set is marked by tenant: $\mathcal{T}$ is the tenant set and $t(x)\in\mathcal{T}$ each point's tenant.
A query has tenant scope $T(q)\subseteq\mathcal{T}$ (single-tenant when $|T(q)|=1$); its effective candidate set is
$\mathcal{X}_p\cap\{x:t(x)\in T(q)\}$. Each tenant $\tau$ carries a service-level objective
$\mathrm{SLO}(\tau)=(\rho_\tau,\ell_\tau)$ (minimum recall $\rho_\tau$, maximum latency $\ell_\tau$) and a
\emph{max-staleness ceiling} $S_\tau$ that the scheduler (\S\ref{sec:system}) honors.

\subsection*{Graph index}
$G=(V,E)$ is a directed graph over $V\subseteq\mathcal{X}$ built by Vamana (out-degree cap $R$, candidate width
$L$, prune slack $\alpha>1$, pool cap \texttt{maxc}); $\mathrm{Search}(G,q,\mathit{ef})$ returns an approximate
top-$k$ using exploration budget $\mathit{ef}$. \textbf{The disk graph is a single \emph{label-blind} Vamana:}
predicates and tenant labels are query-time post-filter tags and are never used in construction. This fact is
central to both the theorem's additivity and topology preservation (\S\ref{sec:theorem},\S\ref{sec:system}).

\subsection*{Mutation streams}
Updates arrive as a marked point process $M=\{(s_i,u_i)\}$ with $u_i\in\{\mathrm{ins}(x),\mathrm{del}(x),
\mathrm{mig}(x,\tau\!\to\!\tau')\}$ and per-class rates $\lambda_{\mathrm{ins}},\lambda_{\mathrm{del}},
\lambda_{\mathrm{mig}}$. The toy setting (\S\ref{sec:theorem}) is Poisson inserts-only
($\lambda_{\mathrm{del}}=\lambda_{\mathrm{mig}}=0$); the general theorem admits all three (migration $=$ a delete
in $\tau$ plus an insert in $\tau'$).

\subsection*{Cache, scheduler, compaction}
A filter-aware, tenant-aware \emph{staging cache} $C$ of capacity $c$ holds pending mutations and is consulted on
the query path, keyed by $(p,T(q))$. A priority-queue \emph{scheduler} orders background graph-repair work by a
composite key $\kappa=\alpha_1\,\mathrm{stale}+\alpha_2\,\widehat{\Delta R}+\alpha_3\,\mathrm{SLA}+
\alpha_4\,\mathrm{mig}$. Compaction fires a merge when cache occupancy reaches $c$ (the theorem's referent for
$c$); the merge integrates the pending batch into $G$ and resets the cache.

\subsection*{Recall as a random variable}
For query $q$ at time $t$, $R_q(t)=|(\mathrm{Search}(G_t,q,\mathit{ef})\cup\mathrm{Cache}_t(q))\cap
\mathrm{GT}(q,t)|/k$. The \emph{per-tenant recall} is $\bar R_\tau(t)=\mathbb{E}_q[R_q(t)\mid T(q)\ni\tau]$, and the
\emph{recall@$k$ loss} is $L_\tau(t)=\rho_\tau-\bar R_\tau(t)$. The object the theorem bounds is the worst-case
\emph{over time} of the query-averaged per-tenant loss on the horizon $[0,T]$,
$L_\tau^*=\sup_{t\in[0,T]}L_\tau(t)$: worst-case in $t$, in expectation over $q$, so the guarantee is on the query
average rather than on any individual query.

\section{The recall-bounded staleness theorem (C1)}
\label{sec:theorem}
We first prove the central kernel in the toy setting (single tenant, no filter, inserts only), then lift it to
the filtered, multi-tenant, deletes/migration setting. The staleness half of the bound is
\emph{semi-empirical / loose-but-correct-form}: its mechanism and functional shape are derived and validated on two
independent axes, while the exponents are calibrated. The filtered half is derived
and parameter-free. We state this status precisely throughout.

\subsection{The toy bound: anti-amortization}
\label{sec:toy}

\begin{definition}[Batch-merge operator $\mathcal{M}_c$]
\label{def:merge}
Let $G_0$ be the disk graph over $N$ points and $B$ a batch of $c$ pending inserts. $\mathcal{M}_c(G_0,B)$ acts in
two phases. \emph{Forward}: each $x\in B$ obtains out-neighbors by greedy search against the \emph{frozen} snapshot
$G_0$ then \textsc{RobustPrune}/$\alpha$-RNG to degree $R$; $G_0$ is not mutated during the batch. \emph{Reconcile}:
each existing node unions its buffered reverse edges and re-prunes (on PQ distances) only if it exceeds $R$. Two
structural consequences follow: \textup{(M1)} the $c$ batch points are \emph{mutually invisible} during selection
(no $x\in B$ takes another batch point as a neighbor, unlike a from-scratch build); \textup{(M2)} hubs flooded by
batch reverse edges are re-pruned lossily on approximate distances.
\end{definition}

\begin{assumption}[Empirical per-merge degradation model, calibrated on this substrate]
\label{ass:permerge}
Integrating one batch of $c$ inserts via $\mathcal{M}_c$ into $G_0$ (over $N$ points) reduces expected disk-graph
recall@$k$, relative to a from-scratch $\alpha$-RNG build on the same final set, by
\begin{equation}
   g(c,N)=\min\!\Big(\,g_0(L)+\Theta\!\big(\min(c^2/N,\,Rc)^{\psi}\big),\ \tfrac{c}{N+c}\,\Big),
\end{equation}
the \textbf{batch term} (M1, super-linear in $c$; $R$-cap-limited, lifted near-linearly $\psi\approx1$ to recall)
plus a small budget-soft \textbf{$V_0$ floor} $g_0(L)$ (M2), capped by the batch fraction.

\emph{Calibration (full fits, sample sizes and diagnostics in \S\ref{sec:eval-accum}).} The exponent is fit on two
independent axes that must not be conflated. On the \textbf{topology} axis (counting missing intra-batch edges) it
is $\theta{=}1.46$ (95\% CI $[1.33,1.59]$), with $b{=}0.50\pm0.04$; on the \textbf{recall} axis (the one the
theorem needs) $a_{\mathrm{recall}}{=}1.28$ (95\% CI $[1.03,1.52]$) at $N_0{=}100$K, and a $\mu$-matched three-base
fit gives $a_{\mathrm{recall}}{=}1.45$. Both place the exponent above $1$ (super-linear) while individual CIs are
wide, so the design rests on the sign, not the value (\S\ref{sec:toy}). The fit is valid over
$0.05\le c/N\le0.45$; below $c/N\approx0.02$ the per-merge loss falls under the $0.02$ pp reporting floor and is
excluded from every fit, so no claim is made for $c/N<0.05$.
\end{assumption}

\begin{theorem}[Toy bound: anti-amortization]
\label{thm:toy}
Under Poisson inserts-only arrivals and Assumption~\ref{ass:permerge}, over horizon $T$ with
$M=\lceil\lambda_{\mathrm{ins}}T/c\rceil$ merges whose per-merge losses assemble additively as an \emph{upper
bound} (measurement finds the assembly subadditive, $\eta<1$, so $M\!\cdot\!g$ over-states $L^*$;
\S\ref{sec:eval-accum}), and $g(c,N)=\Theta(c^{a}N^{-b})$ on the calibrated range $0.05\le c/N\le0.45$,
\begin{equation}
\begin{gathered}
   L^*(c,T)\ \le\ O\!\big(\lambda_{\mathrm{ins}}T\,c^{\,a-1}\,N_0^{-b}\big),\\
   a-1\in[0.03,\,0.52]\ \text{(95\% CI)},\ \ \hat{a}-1\approx0.28,
\end{gathered}
\end{equation}
so this upper bound is \emph{non-decreasing} in the merge size $c$ (strictly increasing on the point estimate) and
$L^*(c)$ itself is measured non-decreasing directly (\S\ref{sec:eval-accum}): the naive
``batch bigger to amortize'' ($L^*\propto1/c$) is wrong in direction. Merge-induced degradation is
\emph{irreducible by batching}, and each merge's damage is capped by $g\le c/(N+c)$, binding only as $c\to N$.
\end{theorem}

\paragraph{Scope of the toy bound.} The central claim is the \emph{sign}: $a\ge1$, i.e.\ batching does not
amortize. That sign forbids re-sizing merges as a recall lever, and it is supported by three fits on independent
data (topology $\theta{=}1.46$, recall $a{=}1.28$ at fixed $N_0$, $\mu$-matched recall $a{=}1.45$). The
\emph{magnitude} is looser ($a-1$ could be as small as $0.03$), so we rest the design on the sign, not the
strength, of the super-linearity. Additive assembly is an assumption, and measurement finds it
\emph{conservative}, with twelve sequential merges at $c{=}5000$, $N_0{=}100$K losing $0.38$--$0.46$ pp cumulatively
against $M\!\cdot\!g=1.04$ pp predicted (accumulation efficiency $\eta\in[0.37,0.44]$); $M\!\cdot\!g$ thus over-states
$L^*$ in the conservative direction for an upper bound, though the constant is loose by $2.3$--$2.7\times$. The
$c$-dependence of $L^*$ at horizon scale is instead \emph{measured directly} (\S\ref{sec:eval-accum}(c)), where the
twelve-merge run repeated across $c\in\{1000,\dots,20000\}$ at a fixed $60$K-insert horizon gives terminal recall
monotonically decreasing in $c$, i.e.\ $L^*(c)$ monotonically \emph{increasing} ($0.42$ pp over a $20\times$ range),
so anti-amortization at horizon scale is a measurement rather than an extrapolation. The same five points also
corroborate the \emph{form}. With $a-1=0.28$ from the independent single-merge fit,
$L^*(c)-L^*(c_{\min})=A(c^{\,a-1}-c_{\min}^{\,a-1})$ holds with $A$ constant to $\pm7\%$, whereas the $c^{\,a}$
construction needs $A$ to move $4.5\times$, so the $1/c$ merge-count factor is measured, not assumed. The $c$-trend
of $\eta$ is reference-dependent and not pinned without a measured from-scratch $160$K anchor, but neither the
sub-additivity nor the measured \emph{sign} of $L^*(c)$ depends on that constant.

The consequence is the design principle of \S\ref{sec:system}: degradation cannot be reduced by re-sizing merges,
only \emph{re-prioritized}. Two mechanisms are isolated (Def.~\ref{def:merge}), M1 mutual invisibility (the
super-linear driver, dominating $c\gtrsim0.1N$) and a small $V_0$ hub-perturbation floor M2 (dominating
$c\lesssim0.1N$). A high-probability version also holds, since the $R$-cap gives $O(R^2)$ bounded influence per insert, so
McDiarmid on the topology divergence and Bernstein on the (sparse, bounded) query loss give
$L^*\le\mathbb{E}[L^*](1+o(1))$ (proof in the supplementary material). Finally, small $c$ (recall-optimal) trades against
tail latency (latency-optimal at large $c$: each merge inflates the tail for $\approx c/d$ seconds), the tension
the scheduler navigates.

\paragraph{The $V_0$ floor does not create an amortization regime.} A natural objection: the floor $g_0(L)$ is
$c$-independent, so paid once per merge it would contribute $M g_0=\lambda_{\mathrm{ins}}T\,g_0/c$, \emph{decreasing}
in $c$; adding it to the increasing batch term would produce an interior recall-optimal $c^\star$ and restore
amortization at small $c$. Measurement settles it: over twelve merges the $V_0$ component \emph{decays}
monotonically ($4.80\to3.43\to2.51\to1.38$ pp at $m{=}1,4,8,12$) rather than staying constant, and the cumulative
loss is $\ll M g_0$. The floor is a saturating transient contributing $O(1)$ rather than $\Theta(M)$, so it
generates no $1/c$ term and $L^*(c)$ has no interior optimum on the measured range. (The decomposition is
scratch-free and not tie-aware, so its absolute magnitudes are confounded, but the conclusion rests only on the sign
of the trend and the decomposition-free check $L_{\mathrm{cum}}\ll Mg_0$.)

\paragraph{Instantiating the constants.} The $\Theta(\cdot)$ hides constants we have measured, so the bound can be
evaluated rather than merely asserted. On the calibrated range, with $g$ in recall points,
\begin{equation}
\begin{gathered}
  g(c,N)\;\approx\;0.087\;(c/5000)^{1.28}\;(N/10^5)^{-0.64}\ \text{pp},\\
  L^*(c,T)\;\lesssim\;\big(\lambda_{\mathrm{ins}}T/c\big)\,g(c,N_0).
\end{gathered}
\end{equation}
Worked example: a tenant taking $10^5$ inserts onto an $N_0{=}10^5$ index at $c{=}10^4$ ($c/N{=}0.1$, inside range)
sees $g=0.21$ pp per merge over $M{=}10$ merges, so $L^*\le2.1$ pp under additive assembly and $\approx0.9$ pp using
the measured $\eta\approx0.4$; the per-merge cap $c/(N+c)=0.091$ is $43\times$ slacker than the realized
$g_{\mathrm{frac}}=0.0021$, so the cap is inactive, as it is everywhere below $c\approx N$.

\subsection{The general theorem (C1-general): filtered, multi-tenant, deletes}
\label{sec:general}

Three assumptions lift the toy bound. The first formalizes local selectivity; the second is the modularity hinge;
the third delimits the predicate class on which the filtered term is a \emph{loss} bound rather than a one-sided
necessity statement.

\begin{assumption}[Local predicate density; the frontier is the post-filtered list]
\label{ass:local}
\emph{$\beta$ is a property of the engine's result contract, not a free constant}: the post-filter can only see
candidates the search returns, so $\beta=1$ whenever the engine surfaces its full width-$L$ list to the filter
(our disk post-filter path, and the measurement of \S\ref{sec:eval-h}, which post-filters a dump of
exactly $L$ candidates), and $\beta<1$ for an engine that truncates to $k$ before filtering. We take $\beta=1$
throughout and carry $\beta$ symbolically only so the law transfers to engines with a different contract.
Filtered recall is governed by the local density $\sigma_{\mathrm{loc}}(p,v)$, not the global $\sigma(p)$: on the
shared graph a width-$L$ greedy search returns a candidate frontier of effective size $L_{\mathrm{eff}}=\beta L$
whose expected in-predicate mass is $\beta L\,\sigma_{\mathrm{loc}}$.
\end{assumption}

\begin{assumption}[Per-merge degradation is filter and tenant-independent]
\label{ass:filterindep}
Because the disk graph is a single label-blind Vamana (\S\ref{sec:model}), $\mathcal{M}_c$ and hence $g(c,N)$
(Assumption~\ref{ass:permerge}) depend only on the label-agnostic vector set, not on the predicate or tenant
partition.
\end{assumption}

\begin{assumption}[Per-query frontier model]
\label{ass:homog}
For a \emph{single} query $q$ with local density $\sigma_{\mathrm{loc}}(q)$, the in-predicate count of its
frontier is $X_q\sim\mathrm{Binom}(\beta L,\sigma_{\mathrm{loc}}(q))$. This constrains one query's neighborhood,
not the query population: $\sigma_{\mathrm{loc}}(\cdot)$ may vary arbitrarily across queries and tenants, and the
tenant-level law is the \emph{average of the per-query law over the query population}
(Lemma~\ref{lem:frontier}(iii)). Substituting a tenant's \emph{mean} local density into the per-query law is a
separate and strictly optimistic step, and spatially-clustered predicates break it
(\S\ref{sec:eval-h}).
\end{assumption}

\noindent Assumption~\ref{ass:filterindep} makes the degradation term \emph{transferable} across the filter and
tenant partition (the same $g(c,N)$ applies regardless of predicate or tenant, i.e.\ modularity), rather than making
the overall bound additive; the additivity of the two loss components is the separate union-bound argument of
Theorem~\ref{thm:general}. It is measured: replaying the scheduler's re-ordered multi-partition merge groupings to
the identical terminal point set changes disk-graph recall by $|\Delta|=0.02$ pp $\le0.67$ pp (\S\ref{sec:eval}),
so merge re-ordering has negligible measured effect on terminal recall.

\begin{lemma}[Post-filter recall is a frontier-mass event]
\label{lem:frontier}
Fix a converged label-blind graph and a query $q$ with local density $\sigma_{\mathrm{loc}}(q)$, and let $X_q$ be
the in-predicate count of its frontier (Assumption~\ref{ass:local}). If the frontier is the true metric top-$\beta
L$, then post-filter recall@$k$ is \emph{exactly} $\min(X_q,k)/k$: the in-predicate members of a true top-$\beta L$
list are precisely the nearest in-predicate points, so the first $\min(X_q,k)$ of them are ground-truth items
(proof in the supplementary material). An approximate search can only \emph{omit} in-predicate ground-truth points
that a true top-$\beta L$ list would have contained, so the identity degrades in one direction: writing
$\delta_{\mathrm{ann}}(L)$ for the expected fraction of a query's in-predicate ground-truth points that lie in the
true top-$\beta L$ but are missed by the graph search,
$\mathbb{E}[\mathrm{recall}]\ \ge\ \mathbb{E}[\min(X_q,k)]/k-\delta_{\mathrm{ann}}(L)$. This is the only direction
Theorem~\ref{thm:general} needs, since it bounds a recall \emph{loss} from above; the reverse
inequality is not claimed. Then:
\begin{enumerate}[topsep=3pt,itemsep=1pt,label=(\roman*)]
\item \emph{(Assumption-free envelope; necessity.)} Without any distributional assumption on $X$,
\begin{equation}
  \mathbb{E}[\min(X,k)]/k\ \le\ \min\!\Big(1,\ \tfrac{\beta L\,\sigma_{\mathrm{loc}}}{k}\Big),
\end{equation}
so $L\ge k/(\beta\sigma_{\mathrm{loc}})$ is \emph{necessary} for recall $\to1$: recovery scale
$L=\Theta(k/\sigma_{\mathrm{loc}})$.
\item \emph{(Exact form under Assumption~\ref{ass:homog}.)} With
$X\sim\mathrm{Binom}(\beta L,\sigma_{\mathrm{loc}})$, expected post-filter recall equals
\begin{equation}
  h(\sigma_{\mathrm{loc}},L,k)\ =\ \frac1k\sum_{i=1}^{k}\Pr\!\big[\mathrm{Binom}(\beta L,\sigma_{\mathrm{loc}})\ge i\big],
\end{equation}
which $\to1$ when $\beta L\sigma_{\mathrm{loc}}\gg k$ and $\to\beta L\sigma_{\mathrm{loc}}/k$ when
$\beta L\sigma_{\mathrm{loc}}\ll k$. The single-term tail
$h_{\mathrm{tail}}=\Pr[\mathrm{Binom}(\beta L,\sigma_{\mathrm{loc}})\ge k]$ is the $i{=}k$ summand alone, so
$h_{\mathrm{tail}}\le h\le\min(1,\beta L\sigma_{\mathrm{loc}}/k)$: a tail-probability model is a valid but
order-of-magnitude-loose lower bound on recall (\S\ref{sec:eval-h}).
\item \emph{(Tenant law = query average; substituting the mean is optimistic.)} A tenant's expected filtered
recall is the \emph{average of the per-query law over its query population},
\begin{equation}
  \bar h_\tau(L,k)\ =\ \mathbb{E}_{q\mid\tau}\big[h(\sigma_{\mathrm{loc}}(q),L,k)\big],
\end{equation}
and $p\mapsto\mathbb{E}[\min(\mathrm{Binom}(n,p),k)]$ is concave in $p$, so by Jensen
\begin{equation}
  \bar h_\tau(L,k)\ \le\ h\big(\mathbb{E}_{q\mid\tau}[\sigma_{\mathrm{loc}}(q)],L,k\big):
\end{equation}
plugging a tenant's \emph{mean} local density into the per-query law over-predicts its recall, with a gap that
grows with the dispersion of $\sigma_{\mathrm{loc}}$ across queries and vanishes where $h$ is linear (small $x$) or
saturated (large $x$). Diffuse predicates have near-zero dispersion and are unaffected; spatially-clustered
predicates are not, and \S\ref{sec:eval-h} exhibits the predicted curvature signature.
\item \emph{(Ordering.)} Collecting (i)--(iii),
$h_{\mathrm{tail}}\le\bar h_\tau\le h(\bar\sigma_{\mathrm{loc}}^{(\tau)},L,k)\le
\min(1,\beta L\bar\sigma^{(\tau)}_{\mathrm{loc}}/k)$, so only the leftmost quantity is a lower bound on recall
without further assumptions, and only the rightmost is an upper bound on recall without any.
\end{enumerate}
\end{lemma}

\emph{Status (\S\ref{sec:eval-h}):} with $\beta{=}1$ fixed by the engine contract (Assumption~\ref{ass:local}) and
\emph{no fitted constant of any kind}, part~(ii) matches all $35$ diffuse cells to a mean absolute residual of
$0.2$ pp (max $1.2$ pp, two-sided), and predicts the separately-measured recovery constant
$L_{95}\approx1.25\,k/\sigma_{\mathrm{loc}}$ because $h$ crosses $0.95$ at $L\sigma_{\mathrm{loc}}\approx1.25k$.
On the clustered controls the residual is \emph{uniformly one-signed} (all $15$ cells over-predict, up to
$+26$ pp) and peaks in the knee $x\approx1$--$1.5$, decaying to $\approx1$ pp at both ends. Two distinct clustering
effects are at work. One is \emph{density variation across queries}, where the dispersion of
$\sigma_{\mathrm{loc}}(q)$ within a tenant makes the mean-substituted $h(\bar\sigma_{\mathrm{loc}},L,k)$ optimistic
by Jensen (Lemma~\ref{lem:frontier}(iii)); this is the curvature signature the $+26$ pp deficit exhibits, not a
failure of the frontier-mass form. The other is \emph{within-frontier correlation}, where spatially-clustered
in-predicate points are not the independent draws Assumption~\ref{ass:homog} posits, so the Binomial
\emph{under}-counts the in-predicate frontier mass; this separate relaxation of the independence assumption acts in
the \emph{conservative} direction, and is why the query-averaged $\bar h_\tau$ remains a lower bound on measured
recall for clustered predicates (\S\ref{sec:eval-h}) rather than an exact predictor.

\begin{theorem}[C1-general: per-tenant recall-bounded staleness]
\label{thm:general}
Under Assumptions~\ref{ass:permerge},~\ref{ass:local},~\ref{ass:filterindep},~\ref{ass:homog} with per-tenant
insert rate $\lambda^{(\tau)}_{\mathrm{ins}}$, local density $\sigma^{(\tau)}_{\mathrm{loc}}$, SLA staleness
ceiling $S_\tau$, and deletes/migration permitted, the worst-case (in $t$) per-tenant recall@$k$ loss decomposes
additively:
\begin{equation}
\begin{aligned}
   L_\tau^*(T)\ \le\ &\underbrace{\kappa_g\,N_\tau^{\mathrm{stale}}\,c^{\,a-1}N_0^{-b}+g_0^{\mathrm{del}}(L,\lambda_{\mathrm{del}})}_{\text{staleness/degradation (filter-independent, Assumption~\ref{ass:filterindep})}}\\[2pt]
   &+\ \underbrace{\big(1-\bar h_\tau(L,k)\big)+\delta_{\mathrm{ann}}(L)}_{\text{filtered navigability (Lemma~\ref{lem:frontier}(ii),(iii))}},
\end{aligned}
\end{equation}
written with an explicit prefactor $\kappa_g$ rather than a $\Theta(\cdot)$, since an inequality whose right-hand
side names an asymptotic class is not a statement one can check. On the calibrated range,
$\kappa_g\approx2.5\times10^{-3}$ pp (fixed by the anchor $g(c{=}5000,N_0{=}10^5)=0.087$ pp with
$a{=}1.28$, $b{=}0.64$); its uncertainty is that of $(a,b)$ and it over-states the assembled loss by
$1/\eta\approx2.3$--$2.7\times$ (\S\ref{sec:eval-accum}), so the inequality holds with slack in the measured regime.
The sum is an upper bound by subadditivity, not an approximation: a ground-truth item is missed if it is
stale-and-uncached \emph{or} absent from the post-filtered frontier, and $\Pr[A\cup B]\le\Pr[A]+\Pr[B]$ over those
two events per item (proof in the supplementary material); the slack is exactly the overlap $\Pr[A\cap B]$, which is why the
statement is an inequality even though Lemma~\ref{lem:frontier}(ii) is an equality in expectation. Here
$\bar h_\tau=\mathbb{E}_{q\mid\tau}[h(\sigma_{\mathrm{loc}}(q),L,k)]$ is the query-averaged filtered term and
$\delta_{\mathrm{ann}}(L)$ is the measurable ANN deficit of the base graph at width $L$.
Here $N_\tau^{\mathrm{stale}}\le\lambda^{(\tau)}_{\mathrm{ins}}(S_\tau+W)$ is \emph{cadence-bounded by the SLA
ceiling}: the scheduler force-drains $\tau$ once its staleness reaches $S_\tau$, but only tests the ceiling at a
merge trigger, so the bound carries one trigger window $W$ of overshoot ($W{=}0$ recovers the idealized
$\lambda^{(\tau)}_{\mathrm{ins}}S_\tau$ and requires an early-merge mechanism not implemented here,
\S\ref{sec:scheduler}). $g_0^{\mathrm{del}}$ is the
delete/tombstone contribution to the $V_0$ floor. The two terms are governed by disjoint parameter sets
($c,N_0,d,S_\tau,\lambda_{\mathrm{del}}$ vs.\ $\sigma^{(\tau)}_{\mathrm{loc}},L,k$); anti-amortization
(Theorem~\ref{thm:toy}) carries into the first term unchanged. The filtered term must be evaluated as the query
average $\bar h_\tau$; substituting the tenant's mean local density gives
$1-h(\bar\sigma^{(\tau)}_{\mathrm{loc}},L,k)$, which by Lemma~\ref{lem:frontier}(iii) \emph{under}-states the loss
and is therefore not a bound, the gap reaching up to $26$ pp for spatially-clustered predicates
(\S\ref{sec:eval-h}). Operationally, a tenant's $L$ must be provisioned from the \emph{distribution} of
$\sigma_{\mathrm{loc}}$ over its queries (a low quantile) rather than from its mean.
\end{theorem}

\begin{remark}[Scope of the bound]
\label{rem:status}
The staleness term is loose-but-correct-form and inherits the toy calibration ($a,b$ calibrated). The filtered term
is now \emph{derived}: Lemma~\ref{lem:frontier}(ii) has no fitted constant at all ($\beta{=}1$ is forced by the
engine's result contract rather than calibrated), and it predicts rather than fits the recovery constant $1.25$.
One residual is carried explicitly instead of being absorbed into a constant: the ANN deficit
$\delta_{\mathrm{ann}}(L)$, which is measurable and not yet measured. The query-dispersion gap of
Lemma~\ref{lem:frontier}(iii) has instead been evaluated on the archived per-query $\sigma_{\mathrm{loc}}$: the
query average $\bar h_\tau$ is a lower bound on delivered recall that is \emph{empirically conservative to the reported precision} (measured $\ge\bar h_\tau$ in
$14$ of $15$ clustered cells; the sole exception under-runs by $0.09$ pp, within reporting precision, \S\ref{sec:eval-h}), loose by up to $14$ pp at low $\sigma_{\mathrm{loc}}$/small $L$. The $k$-dependence of the law is a sharp untested prediction: $h$ is a function of
$(\beta L\sigma_{\mathrm{loc}},k)$ and \emph{not} of $x=\sigma_{\mathrm{loc}}L/k$ alone: at fixed $x{=}1$ it
predicts $63.2/87.4/96.0$\% recall for $k{=}1/10/100$, whereas every measurement here is at $k{=}10$. The deletes
floor
$g_0^{\mathrm{del}}$ is parameterized but not yet characterized, and a first-principles derivation of the
degradation exponents from $\alpha$-RNG occlusion geometry is open; both remain open items. On the
shared label-blind graph, filtered recall degrades with $\sigma_{\mathrm{loc}}$: the theorem's \emph{content}, not
an unmodeled loss; a per-tenant-navigable structure removes it but cannot stream and carries no bound.

In sum, three of Theorem~\ref{thm:general}'s four parts hold unconditionally (the additive assembly by
subadditivity, the derived parameter-free filtered term, and the named measurable ANN deficit), while the
staleness term is a \emph{calibrated scaling law} (its prefactor and exponents fit on one substrate, dataset family,
and $c/N$ range). The theorem is thus a proved decomposition of which one component is empirical: weaker than a
distribution-free $(\epsilon,\delta)$ guarantee, stronger than a recall estimator that predicts without
constraining. A first-principles derivation of the degradation exponent would replace the calibration with a
constant; we state it as open.
\end{remark}

\section{System: cache (C2) and scheduler (C3)}
\label{sec:system}

The two engineering contributions instantiate the theorem. C2 (the staging cache) absorbs within-cycle staleness
on the query path so the loss the theorem bounds is the disk-graph term, not cache coverage; C3 (the scheduler)
re-prioritizes background repair without ever re-sizing a merge, which is the only lever anti-amortization leaves.
We prototype both on a FreshDiskANN fork with an \texttt{io\_uring} backend. \emph{All numbers below are
measured on this prototype unless marked otherwise.}

\subsection{Substrate: one label-blind Vamana}
\label{sec:substrate}
The disk index is a \emph{single} Vamana graph over the label-agnostic vector set; predicates and tenant labels
are query-time post-filter tags and never enter graph construction or the merge operator $\mathcal{M}_c$. Two
properties follow, both central to \S\ref{sec:theorem}. First, the per-merge degradation $g(c,N)$ is
\emph{filter and tenant-independent} (Assumption~\ref{ass:filterindep}), the additivity hinge of
Theorem~\ref{thm:general}. Second, a partition boundary can never induce topology degradation, because the graph
never sees one; the scheduler's re-ordering only changes which vectors co-merge and in what order. The prototype
inherits the baseline's merge exactly, and by design rather than by tuning: at every trigger all selected
partitions are folded into \emph{one} combined in-memory index and handed to \emph{one} merge of size $c$, so the
merge operator receives the same point set, at the same cadence, with the same $\mathrm{Merge\_Size}$ as the
unmodified baseline. Partitioning is therefore invisible to $\mathcal{M}_c$ as a matter of code path, which is why
the bound carries. We check that construction rather than infer it: with a single predicate and tenant the full C2+C3 prototype
reproduces baseline post-merge recall@10 to $\delta{=}0.000$ at three successive checkpoints
($99.286/99.145/99.018$, brute-force $K{=}100$ ground truth), with the ceiling-disabled, ceiling-never-firing, and
FIFO legs byte-identical.

\subsection{C2: a filter-aware, tenant-aware staging cache}
\label{sec:cache}
FreshDiskANN stages pending inserts in two in-memory buffers double-buffered against the disk merge. We re-key each
buffer to a hash map over \emph{partitions} $(\textit{predicate},\textit{tenant})$ and consult it
on the query path, routing a query to its own partition and label-post-filtering disk results. The cache
therefore serves freshly-inserted, same-tenant points immediately, which is what makes filtering on the query
path viable and is essential for filtered queries, since disabling the query-path keying costs $72$--$86$ pp of
per-label recall@10 at live cache occupancy ($93.8/78.9/81.1$ keyed vs.\ $7.8/7.3/7.6$ unkeyed, at comparable p99,
with \emph{identical} probed staging occupancy in both legs so the staging state is held fixed). This ablation
toggles more than a cache-probe comparison. A single query key drives
both the cache probe and the disk post-filter, so the unkeyed leg is not a filter-blind cache alongside a correct
disk path, since it pins every query's key to one default partition, which simultaneously blinds the cache probe and
points the disk post-filter at the wrong label for every tenant but one. The residual $7.5\%$ is accordingly close
to arithmetic, approximately the one correctly-keyed tenant's share of the query mix times its recall. What
the ablation establishes is that \emph{the query-path keying as a whole} is central; it does not measure a
cache-probe-only lift. Isolating the probe would require holding the disk post-filter on the true label while
blinding only the cache, which the current knobs do not expose; the tighter comparator (a filter-blind cache that
post-filters its own contents) is therefore not measured. The cache changes only
\emph{what is staged and how queries route}, not the merge cadence; at a merge, all selected partitions are folded
into \emph{one} combined in-memory index and drained as a \emph{single} merge of size $c$ (the combined-index
single-merge drain), identical in topology to the baseline (merge-order parity $\delta{=}0.000$). This is what
lets the bound carry to the prototype unchanged. Unlike read-side or filter-blind caches over vector
indexes (Ubis, MicroNN, QVCache), ours is jointly filter and tenant-keyed, on the query path, and is the exact
structure the theorem binds.

\subsection{C3: an SLA-aware, select-not-resize scheduler}
\label{sec:scheduler}
The scheduler \textbf{never re-sizes a merge}; every fired merge is one combined fixed-size-$c$ drain (bound-safe by
construction, \S\ref{sec:substrate}), and it wins by \emph{selecting which pending work fills each merge and when}.
The rule follows from Theorem~\ref{thm:toy}, by which re-sizing is not a recall lever in the first place, so there
is nothing to gain, and from holding the merge at the baseline's shape, which keeps the bound applicable without
re-deriving it per configuration. An implementation observation that draining $N$ partitions as
$N$ separate merges reads much worse ($84.9/76.4/70.1$) than the one-combined-merge design is \emph{not} used as
evidence, since it holds nothing fixed, and the design rule rests on the two reasons above.

It orders background repair by a composite key
\begin{equation}
   \kappa(w,t)=\alpha_1\,\mathrm{stale}(w,t)+\alpha_2\,\widehat{\Delta R}(w)+\alpha_3\,\mathrm{SLA}(w)+\alpha_4\,\mathrm{mig}(w),
\end{equation}
over staleness, predicted recall impact, tightest tenant SLA, and migration cost, each min--max normalized across
the pending set at drain time. \emph{Exactly one of these terms is exercised in every measurement we report.} The $\widehat{\Delta R}$ term is implemented as an occupancy proxy behind
a default-off flag and is disabled ($\alpha_2{=}0$) throughout, and no result depends on it. The migration term is identically zero in this prototype.
The SLA term is inert whenever tenants share one SLO, since a term with a single distinct value normalizes to zero,
and our churn legs additionally set $\alpha_3{=}0$ so that ceiling enforcement is the only variable. So $\kappa$
reduces to staleness ordering (oldest-first) in all reported runs; deferred units reload as the oldest across the
driver's per-merge restart and drain first. The four-term form is the interface the
scheduler exposes, not a mechanism this paper validates; the measured win below comes from the ceiling.

On top of $\kappa$, a \textbf{per-tenant max-staleness ceiling} $S_\tau$ force-drains a tenant whose staleness has
reached $S_\tau$. This is a \emph{prioritization with a cadence-bounded overshoot}, and the distinction matters for
Theorem~\ref{thm:general}. The built scheduler tests ceilings only at a merge trigger, with no early-merge on
crossing, so a tenant that crosses $S_\tau$ mid-window stays stale until the next trigger. The guarantee we
actually deliver is
\begin{equation}
   N^{\mathrm{stale}}_\tau\;\le\;\lambda^{(\tau)}_{\mathrm{ins}}\big(S_\tau+W\big),
\end{equation}
where $W$ is the trigger window (the high-water mark). The theorem's $\lambda^{(\tau)}_{\mathrm{ins}}S_\tau$ cap is
thus the $W\!\to\!0$ idealization; the prototype attains it up to one window, and we set $W=S_\tau^{\min}$ to make
that overshoot as small as the design admits without an early-merge mechanism. The effect is directly visible in
our runs, where a tight tenant's measured staleness reaches $1900$ against a $1000$ ceiling before the force-drain
resets it, and in an earlier configuration (even with $W$ set equal to $S_\tau$, the tightest the built cadence
allows) the tight tier was still violated at one checkpoint on \emph{both} legs for exactly this reason. Closing
the gap would need an early-merge trigger.
FIFO/threshold-triggered compaction (all prior streaming systems) has no per-tenant lever at all;
\S\ref{sec:eval-churn} shows the ceiling is what drives tight-tenant violations to zero under churn.

\subsection{Deletes and tenancy migration}
\label{sec:deletes}
Deletes are tombstoned in the staging partitions and evicted at the combined merge; a tenancy migration is a delete
in $\tau$ plus an insert in $\tau'$. We extend the drain to evict tombstones from \emph{every} partition (not only
a default one), verified to conserve points (zero silently-dropped) and to leave deleted ids absent from
post-merge results while survivors remain. This is what admits the deletes/migration term of
Theorem~\ref{thm:general}; the term's precise floor contribution $g_0^{\mathrm{del}}$ is not characterized here.

\section{Evaluation}
\label{sec:eval}

\paragraph{Setup.} Prototype on a FreshDiskANN fork, \texttt{io\_uring}; SIFT at $10$K (robust local
scale) and $100$K (mid-scale), plus a filtered YFCC slice; Vamana $R{=}32$, build/search $L{=}75$, beam $4$,
$k{=}10$. Every filtered/multi-tenant recall uses \emph{per-label} ground truth, built as the true top-$k$ among
same-label points by brute force. Unfiltered ground truth would floor label-filtered recall at $\approx1/L_{\mathrm{pred}}$; on one label at
$2{,}500$ queries the same run reads $24.4$ against unfiltered ground truth versus $86.6$ against per-label ground
truth, a $+62$ pp gap that brackets the $1/6$ floor. Every reference used here is brute-force $K{=}100$ in the
two-block format the loader expects. Disk-graph numbers are cache-empty; we state the engine and full config on each. Determinism (insert-order invariance) means exponent uncertainty is the log-log
fit CI, not run replication.

\subsection{Merge-order re-ordering has negligible effect on terminal recall}
\label{sec:eval-topo}
The general theorem's additivity rests on the scheduler not degrading the terminal graph
(Assumption~\ref{ass:filterindep}). We replay the enforce-ordered and FIFO-ordered multi-partition merge groupings
to the \emph{identical} $18{,}000$-point terminal and read cache-empty recall@10 under an identical
search over the full $10{,}000$-query set: $\Delta=\text{recall(enforce)}-\text{recall(FIFO)}=54.58-54.60=-0.02$
pp, i.e.\ $|\Delta|=0.02$ pp $\le0.67$ pp (Table~\ref{tab:topo}; also Fig.~S1). Re-ordering therefore leaves terminal recall
unchanged to within $0.02$ pp, the order-invariance the label-blind design predicts (Assumption~\ref{ass:filterindep}).
The instrument measures recall, so the claim is functional equivalence at the recall level.

The instrument's absolute value ($54.6$) sits far below every other recall we report ($\approx99.7$) for a benign
reason. To obtain a single-method $\Delta$ we searched both terminals with one standalone tool that scores all
$18{,}000$ live points against ground truth built for the first $8{,}000$, so about half of each query's true
neighbours lie outside the scored reference. This depresses both legs \emph{equally}, cancelling in $\Delta$ though
not in the level (the aligned in-process path reads $99.911$ on the enforce leg but hangs on the other leg on
this box, so the aligned-absolute $\Delta$ is deferred). The $0.02$ pp is a real graph difference rather than noise, since both
indexes are deterministic and insert-order replication is byte-identical across seeds ($\sigma=0.000$ pp). The
$0.67$ pp acceptance band is a pre-registered prior threshold, not a power calculation; it comes from a separate
single-partition deferred-vs-FIFO experiment ($97.0$ vs.\ $97.67$ at $80$K) that over-estimates the compositional
effect because its terminal drain was one large merge (carrying the merge-size effect of Theorem~\ref{thm:toy}); we
claim only that the measured $\Delta$ is two orders of magnitude inside it, with no calibrated power to exclude
differences between $0.02$ and $0.67$ pp.

\begin{table}[tbp]\centering\scriptsize
\caption{Terminal-recall $\Delta$ under merge re-ordering (identical $18$K terminals, same-method relative).}
\label{tab:topo}
\begin{tabular}{lcc}
\toprule
leg & terminal pts & recall@10 (same tool)\\
\midrule
enforce-ordered & 18{,}000 & 54.58\\
FIFO-ordered & 18{,}000 & 54.60\\
\bottomrule
\end{tabular}\\[2pt]
{\footnotesize $|\Delta|=0.02$ pp, inside the $\le0.67$ pp band.}
\end{table}

\subsection{Anti-amortization: the per-merge law, and how it assembles over a horizon}
\label{sec:eval-accum}
This subsection carries the empirical weight of Theorem~\ref{thm:toy}, whose two factors are measured separately:
the \emph{per-merge} law $g(c,N)$ (measured directly and cleanly), and its \emph{assembly} over $M$ merges
(measured at a single $c$, and sub-additive).

\paragraph{(a) Per-merge law (single-merge isolation, $N_0{=}100$K, $L{=}75$).} We restore a pristine base graph,
insert exactly $c$ fresh points, force one merge, reload with an empty memory index, and read cache-empty recall@10 over $10$K SIFT queries against ground truth for the identical live set; the reference
$R_{\mathrm{scratch}}$ is a from-scratch Vamana on the same points with identical parameters. Table~SI (supplementary material)
gives $g=R_{\mathrm{scratch}}-R_{\mathrm{merge}}$. The fitted exponent is $a=1.28$ (95\% CI $[1.03,1.52]$,
$R^2{=}0.99$) after excluding the two cells below the $0.02$ pp floor, confirming a super-linear trend; the local
slopes across successive doublings ($1.17,0.88,1.64,1.44$) straddle $1$, so the trend holds over the range without
monotone super-linearity at every $c$. The batch-fraction cap acts as a structural ceiling rather than a predictive
law, with $g_{\mathrm{frac}}$ sitting $10$--$300\times$ under $c/(N+c)$ throughout. Because $R_{\mathrm{scratch}}$
varies only $0.18$ pp across the sweep ($99.71\to99.53$), the $c$-comparison is not an artifact of a moving reference.

\paragraph{(b) Assembly over a horizon (twelve sequential merges, $c{=}5000$, $N_0{=}100$K).} The theorem multiplies
$g$ by $M$. We test that directly: a streaming inserts-only run with $M{=}12$ forced merges of $c{=}5000$ each,
reading cache-empty recall at every merge boundary against ground truth matched to the live set. The first
merge reproduces the isolation harness to $0.006$ pp ($99.627$ vs.\ $99.633$), validating the two measurements
against each other. Cumulative loss then grows to $0.384$ pp by $m{=}12$ (relative to $m{=}1$; $\approx0.46$ pp
against a from-scratch $160$K reference), against $M\!\cdot\!g=1.04$ pp predicted by additive assembly (Table~SII, supplementary material). The
per-merge increment settles to $\approx0.024$ pp after a four-merge transient, with no downward trend, so the loss
is linear in $M$ in \emph{steady state}, but at a per-merge rate $3.6\times$ below the isolated single-merge $g$.
The mechanism is visible in a per-epoch decomposition of the missed top-$k$: the $V_0$ (pre-existing point)
component decays with $m$ while the newly-inserted-frontier component grows: later merges damage the recent
frontier rather than re-damaging the base. This carries one favourable implication and one cost. Favourably, the
$c$-independent $V_0$ floor saturates instead of accumulating, removing the $Mg_0\propto1/c$ term and with it any
interior recall-optimum in $c$ (\S\ref{sec:toy}). The cost is that $M\!\cdot\!g$ over-states the horizon loss by
$2.3$--$2.7\times$ at this $c$, so the additive bound is conservative in this regime, and whether $\eta$ is
$c$-dependent (which could in principle flatten $L^*(c)$) is settled by part (c). The decomposition is scratch-free and not tie-aware, so its
absolute magnitudes are confounded, and the conclusions above use only its sign and the decomposition-free
comparison $L_{\mathrm{cum}}\ll Mg_0$.

\paragraph{(c) The horizon law $L^*(c)$, measured directly.} We repeat the twelve-merge protocol across
$c\in\{1000,2500,5000,10000,20000\}$ at a \emph{fixed} $60$K-insert horizon (merge counts $M=60,24,12,6,3$; same
base graph, seed, and cache-empty protocol; the $c{=}5000$ leg reproduces part (b) to the decimal). Every
$c$ reaches the same $160$K live set scored against the same ground truth, so the sign of $L^*(c)$ is the sign of
the terminal-recall trend and needs no from-scratch reference. Terminal recall@10 \emph{decreases monotonically}
in $c$ ($99.436, 99.343, 99.243, 99.168, 99.013$ at $c=1$K$/2.5$K$/5$K$/10$K$/20$K), so $L^*(c)$ is monotonically
\emph{increasing} in $c$ (a $0.42$ pp spread over a $20\times$ range, five points, clean). \textbf{Anti-amortization
therefore holds at horizon scale as a direct measurement, not an extrapolation.} In particular the $c{=}1000$ and
$2500$ legs lie below the $0.05\le c/N\le0.45$ calibration range of Assumption~\ref{ass:permerge} and enter only
this direct monotonicity measurement, never the fitted power-law. Fewer-but-larger merges leave a
worse final graph than more-but-smaller ones. The accumulation efficiency $\eta(c)=L^*/(Mg)$ is below $1$ at every
above-floor $c$ (assembly is sub-additive, since per-merge damage saturates against the $V_0$ floor rather than
stacking, which is why $M\!\cdot\!g$ over-states $L^*$). Its \emph{trend} in $c$ is reference-dependent, so the
direction it appears to take is a property of the convention. Read against each run's own first merge, $\eta=0.37,0.29,0.25$ (decreasing) at
$c=5$K$/10$K$/20$K; read against a from-scratch $160$K graph (the convention the theorem's $g$ uses), the same
$L^*$ values give $\approx0.44,0.46,0.64$ (increasing). The two differ only in the reference recall, and the
from-scratch anchor $R_{\mathrm{scratch}}(160\text{K})$ is estimated rather than measured, so the sign of
$d\eta/dc$ is not pinned. What \emph{is} pinned is that $\eta<1$ and that $L^*(c)$ itself increases; the loose
constant bends the magnitude but does not flatten or reverse the measured sign. ($c\le2000$ has
sub-floor single-merge $g$, so its $\eta$ is ill-defined; its $L^*$ points need no $g$ and are valid.) This also
corrects a takeaway from the single-$c$ run, since recall is \emph{not} monotone-better in larger $c$ but penalises
it, so recall joins staleness and staging-RSS in favouring bounded $c$, strengthening select-not-resize.

\paragraph{The sweep also corroborates the functional form, on independent data.} These five points test more than
a sign. The theorem predicts $L^*(c)=A\,c^{\,a-1}$; read against the $c{=}1000$ leg, the testable form is
$L^*(c)-L^*(c_{\min})=A\,(c^{\,a-1}-c_{\min}^{\,a-1})$ with a single free scale $A$. Fixing $a-1=0.28$ from the
\emph{single-merge} recall fit (a different experiment), the four remaining points give
$A=0.046,0.049,0.043,0.047$ (mean $0.0461$, $\pm7\%$, no drift): the exponent is approximately right in magnitude,
transferred across experiments without refitting. The same construction with $c^{\,a}$ instead of $c^{\,a-1}$
requires $A$ to move $4.5\times$ ($6.0,4.1,2.1,1.4\times10^{-6}$), so the $M=\lambda_{\mathrm{ins}}T/c$ merge-count
factor is doing real work. Three limitations: $A$ absorbs the accumulation efficiency and so is calibrated, not
predicted (only its \emph{constancy} is tested); the fixed-reference construction makes the four residuals
dependent on the $c{=}1000$ leg; and with $a-1$ uncertain to $[0.03,0.52]$ the fit corroborates $0.28$ without
sharpening the CI.

\paragraph{Topology cross-check and the $\psi\approx1$ lift.} The recall-axis exponent is corroborated by an
independent method: node-by-node edge counting on the merged versus from-scratch graphs gives a
missing-intra-batch-edge exponent $\theta=1.46$ (95\% CI $[1.33,1.59]$), matching a recall exponent
$\gamma=1.46\pm0.14$ from a separate three-replicate streaming sweep, the evidence for the near-linear
topology-to-recall lift $\psi\approx1$. The edge count also shows the M1 mechanism at its strongest: \emph{every}
intra-batch edge a from-scratch build would form is absent after a merge (ratio $1.000$ at all five $c$), with
out-degree distributions identical (all nodes at the $R{=}32$ cap), so the divergence is purely in which neighbours
are chosen. Two qualifications: the lift is established only to $\pm10$--$15\%$, and on the $N$ axis the transfer is
weaker ($b{=}0.50\pm0.04$ from topology vs.\ $b_{\mathrm{recall}}{=}0.64$ from the $\mu$-matched recall fit). An
earlier matched-$c$ fit read $b_{\mathrm{recall}}\approx1.3$ and appeared to refute $\psi\approx1$; holding
$\mu=kc/N$ fixed across three bases showed that was a regime-confound (the two bases sat at different $c/N$) and
recovered $a_{\mathrm{recall}}{=}1.45$. We carry the residual $b$ gap ($0.50$ vs.\ $0.64$) as a mild unexplained
steepening rather than absorbing it.

\subsection{The filtered term $h$: recall is the truncated frontier mass, with $\beta{=}1$}
\label{sec:eval-h}
We sweep local density $\sigma_{\mathrm{loc}}$ (diffuse random-balanced labels and spatially-clustered labels at
matched \emph{global} selectivity) against search width $L$, measuring post-filter recall@10 with per-label GT on
one label-blind base graph (Table~\ref{tab:sigmaloc}).

\paragraph{(1) The collapse variable is $x=\sigma_{\mathrm{loc}}L/k$, and it is tight.} Cells with equal
$\sigma_{\mathrm{loc}}L$ but different $(\sigma_{\mathrm{loc}},L)$ agree closely (e.g.\ four independent pairs at
$x{=}1.26$ give $95.42/95.27/95.29/95.40$), and the collapse is scale-invariant $10$K$\leftrightarrow100$K.
Local, not global, selectivity governs recall. The clustered slice at global $\sigma=0.042$ (which
\emph{global} density would place at ${\approx}31$) recalls $73.5$ because its measured
$\hat\sigma_{\mathrm{loc}}=0.168$; $\sigma_{\mathrm{loc}}$ over-predicts it by a comparable margin (finding 3), so
it is the better of two imperfect predictors rather than a sufficient statistic.

\paragraph{(2) $\beta{=}1$ is structural, and the law then has no free parameter.} The measurement post-filters the
search's returned candidate list, of length exactly $L$, and takes the top-$k$ of the survivors; a frontier of
$1.66L$ candidates cannot be filtered out of a list of $L$, so $\beta{=}1$ is forced by the pipeline rather than
fitted (Assumption~\ref{ass:local}). The data agrees independently: in the unsaturated regime measured recall
equals $\sigma_{\mathrm{loc}}L/k$ with unit slope (at $x=0.078,0.156,0.313,0.629$ the ratio recall$/x$ is
$1.00,1.00,1.00,0.98$), which is the exact small-$x$ limit of Lemma~\ref{lem:frontier}(ii) only when $\beta{=}1$.
With no fitted constant, $h$ matches all $35$ diffuse cells to within $1.2$ pp, sub-pp on $34$ of $35$
(Table~SIII, supplementary material; largest residual $-1.2$ pp at the extreme corner $\sigma_{\mathrm{loc}}{=}0.0103$,
$L{=}1000$), and residuals are two-sided, so the form is not systematically optimistic on uniform predicates.
Independently, $h$ crosses $0.95$ at $\sigma_{\mathrm{loc}}L\approx1.25k$, i.e.\
$L_{95}\approx1.25\,k/\sigma_{\mathrm{loc}}$ with exponent $-1.0$, so the recovery constant is \emph{predicted}
rather than fitted. There is no hard break down to $\sigma_{\mathrm{loc}}=0.01$.

\paragraph{(3) A tail-probability model is the wrong shape.} The single-term tail
$h_{\mathrm{tail}}=\Pr[\mathrm{Binom}(\beta L,\sigma_{\mathrm{loc}})\ge k]$ is a far \emph{sharper} function of $x$
than the data, being the $i{=}k$ summand of the $k$-term average in Lemma~\ref{lem:frontier}(ii) and so
under-predicts recall by $\sim8\times$ at $x{=}0.31$ and by $\sim10^5\times$ at $x{=}0.078$
(Table~SIII, supplementary material). Inflating $\beta$ to $1.66$ rescues the fit only near $x\approx0.63$ and makes the tail
form \emph{over}-predict recall for $x\gtrsim1$ ($99.7$ vs.\ $95.4$ measured at $x{=}1.26$; $96.0$ vs.\ $90.0$ at
$x{=}1.03$), i.e.\ non-conservative exactly where a loss bound needs conservatism. This is why we carry the
truncated mean with $\beta{=}1$ and not a re-fitted tail.

\paragraph{(4) The clustered deficit is one-signed and carries the predicted curvature signature.} All $15$
clustered cells sit \emph{below} $h$ evaluated at their measured mean $\hat\sigma_{\mathrm{loc}}$, never above, and
the deficit is not monotone in $L$: for the $T{=}96$ clustered control it runs $+6.1,+21.7,+26.0,+7.2,+1.1$ pp
across $L=75\ldots1000$, peaking in the knee $x\approx1.4$ and decaying to $\approx1$ pp at both ends. That shape
is what Lemma~\ref{lem:frontier}(iii) requires and what a genuine breakdown of the frontier-mass form would not
produce: the Jensen gap must vanish where $\gamma$ is affine (linear at small $x$, flat at saturation) and peak
where it is most curved. The one-signedness is the correct direction, since dispersion in
$\sigma_{\mathrm{loc}}$ across a tenant's queries can only lower recall. We attribute the deficit to evaluating $h$
at the mean rather than averaging it over queries; the query average
$\bar h_\tau=\mathbb{E}_q[h(\sigma_{\mathrm{loc}}(q),L,k)]$, computable from the per-query $\sigma_{\mathrm{loc}}$ the
sweep records, closes the $26$ pp gap (see the query-average result below).

\paragraph{(5) Tenant count does not set $\sigma_{\mathrm{loc}}$; label geometry does.} This distinction decides
whether the law undermines or supports the multi-tenant thesis, so we state it explicitly. Read naively, the law
looks self-defeating, since more tenants means smaller per-tenant selectivity $\sigma{=}1/T$, hence (via
$L_{95}{=}1.25k/\sigma_{\mathrm{loc}}$) a search width growing linearly in $T$, so the very regime the system
targets would be the regime it serves worst. That reading conflates $\sigma_{\mathrm{loc}}$ with $\sigma$, and the
clustered controls refute it. Random-balanced labels make $\hat\sigma_{\mathrm{loc}}$ equal $1/T$ to within $0.5\%$
at every $T$. They are the \emph{worst case} for locality, since a tenant's points are scattered uniformly through
the space. Spatially-coherent labels, which model workloads in which a tenant's vectors exhibit domain-specific locality,
give $\hat\sigma_{\mathrm{loc}}$ of $0.511/0.168/0.047$ against global $\sigma$ of $0.167/0.042/0.010$ at
$T{=}6/24/96$, a $3$--$5\times$ locality premium that does not decay with $T$. The consequence at matched search
cost is large. At $L{=}75$ the clustered controls recall $97.8/73.6/29.2$ where the diffuse controls at the same
tenant counts recall $95.4/31.4/7.8$, a $2.3\times$ and $3.8\times$ improvement at $T{=}24$ and $T{=}96$, obtained
purely from label geometry and \emph{despite} the clustered cells' own dispersion deficit (finding 4). What the law
penalizes is therefore predicate diffuseness, not multi-tenancy: a thousand tenants with coherent data are cheaper
to serve than twenty-four with data sprayed uniformly. Real workloads need not sit at the clustered extreme;
$\hat\sigma_{\mathrm{loc}}$ is a measurable property of a deployment and the input the law needs.

\paragraph{(6) The joint recall--latency feasibility test.} Because recovery costs
$L_{95}{=}1.25k/\sigma_{\mathrm{loc}}$ and per-query cost grows with $L$, a tenant carrying both a recall floor
$\rho_\tau$ and a latency SLO $\ell_\tau$ is servable on the shared substrate only if
\begin{equation}
   1.25\,k/\hat\sigma_{\mathrm{loc}}^{(\tau)}\;\le\;L_{\max}(\ell_\tau),
\end{equation}
where $L_{\max}(\ell_\tau)$ is the widest search meeting $\ell_\tau$. This is an admission-control and provisioning
test, computable before deployment from a quantity the operator can measure: the practical content of
``recoverable at $\Theta(k/\sigma_{\mathrm{loc}})$''. It has a hard consequence: at $\hat\sigma_{\mathrm{loc}}{=}0.01$
the test demands $L\approx1250$ ($17\times$ our default $L{=}75$), which \S\ref{sec:eval-latency} prices at
${\approx}275$ ms mean, outside any interactive budget, so such tenants cannot meet a $95\%$ recall floor and an
interactive SLO simultaneously on the shared substrate, and a per-tenant-navigable index is the right structure (the
same boundary on which \S\ref{sec:eval-strawmen} places Curator ahead, now a quantitative condition). Mean latency
is linear in $L$ over the same ladder (\S\ref{sec:eval-latency}), and since
the search is label-blind the price of a recall target depends only on the $L$ that $\hat\sigma_{\mathrm{loc}}$
demands, not on the predicate. The test is stated on mean latency and against a candidate SLO ladder, and
\S\ref{sec:eval-latency} measures the latency side directly.

\paragraph{On the strength of the diffuse evidence.} Two limits apply. The diffuse $\hat\sigma_{\mathrm{loc}}$
agrees with $1/T$ to within $0.5\%$ at every $T$, which validates the $\sigma_{\mathrm{loc}}$ instrument but also
means the seven diffuse controls form a single scaling family, so their mutual collapse is largely structural. And
every measurement here is at $k{=}10$, leaving the law's $k$-dependence (Remark~\ref{rem:status}) untested.

\begin{table*}[tbp]\centering\scriptsize
\caption{Post-filter recall@10 vs.\ $(\hat\sigma_{\mathrm{loc}},L)$, SIFT-10K, per-label GT. Global selectivity
$\sigma{=}1/T$ shown for contrast: for spatially-coherent (clustered) tenants $\hat\sigma_{\mathrm{loc}}$ exceeds
$\sigma$ several-fold at every tenant count, and it is $\hat\sigma_{\mathrm{loc}}$ that governs recall.}
\label{tab:sigmaloc}
\begin{tabular}{llcccccc}
\toprule
control ($T$) & $\sigma{=}1/T$ & $\hat\sigma_{\mathrm{loc}}$ & $L{=}75$ & $150$ & $300$ & $600$ & $1000$\\
\midrule
diffuse (6)   & 0.167 & 0.168 & 95.42 & 99.97 & 99.99 & 100.0 & 100.0\\
diffuse (12)  & 0.083 & 0.084 & 61.87 & 95.27 & 99.98 & 100.0 & 100.0\\
diffuse (24)  & 0.042 & 0.042 & 31.36 & 61.54 & 95.29 & 99.99 & 100.0\\
diffuse (48)  & 0.021 & 0.021 & 15.63 & 31.06 & 61.03 & 95.40 & 99.94\\
diffuse (96)  & 0.010 & 0.010 &  7.75 & 15.51 & 30.98 & 61.41 & 89.99\\
\midrule
clustered (6) & 0.167 & 0.511 & 97.78 & 99.47 & 99.84 & 99.94 & 99.97\\
clustered (24)& 0.042 & 0.168 & 73.55 & 92.30 & 98.71 & 99.83 & 99.96\\
clustered (96)& 0.010 & 0.047 & 29.18 & 47.17 & 71.61 & 92.80 & 98.87\\
\bottomrule
\end{tabular}
\end{table*}

\begin{figure}[tbp]\centering\includegraphics[width=0.8\columnwidth]{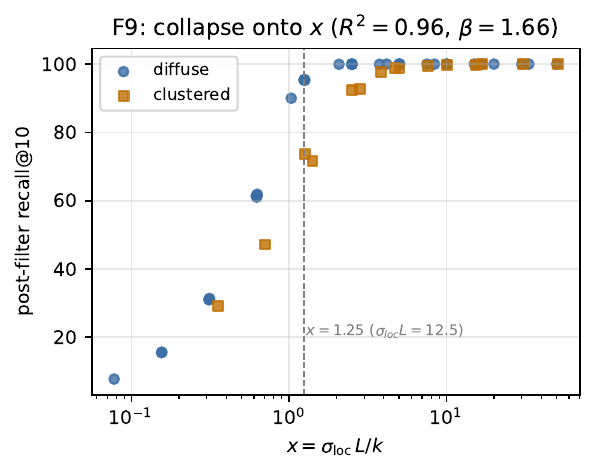}
\caption{Post-filter recall@10 collapses onto $x=\sigma_{\mathrm{loc}}L/k$ with \emph{unit slope} in the
unsaturated regime (diffuse + clustered controls). Overlaid: the truncated-mean law
$h=\frac1k\sum_{i\le k}\Pr[\mathrm{Binom}(L,\sigma_{\mathrm{loc}})\ge i]$ ($\beta{=}1$, no fitted constant),
the Jensen envelope $\min(1,x)$, and the previously-used single-term tail with $\beta{=}1.66$ (too sharp in $x$:
under-predicts for $x{<}0.5$, over-predicts for $x{>}1$). Recovery $L_{95}=1.25\,k/\sigma_{\mathrm{loc}}$ is
predicted by $h$, not fitted. The clustered control sits below the curve, as
Lemma~\ref{lem:frontier}(iii) requires.}\label{fig:F9}\end{figure}

\paragraph{The per-tenant guarantee holds as a query average.} Table~SIII (supplementary material) evaluates $h$ at each
control's \emph{mean} $\hat\sigma_{\mathrm{loc}}$, which Lemma~\ref{lem:frontier}(iii) predicts is optimistic for
dispersed (clustered) predicates, and indeed the clustered residuals there are one-signed and peak in the knee.
Evaluating the term as the theorem actually uses it (the query average
$\bar h_\tau=\mathbb{E}_{q\mid\tau}[h(\sigma_{\mathrm{loc}}(q),L,k)]$, from the per-query $\sigma_{\mathrm{loc}}$
distribution) removes the discrepancy, with measured recall $\ge\bar h_\tau$ in \textbf{14 of 15} clustered cells
(the 15th over by $0.09$ pp, within rounding). So the per-tenant term is a \emph{conservative lower bound} on delivered
recall rather than an equality, since the mean-substitution gap runs both ways (optimistic by up to $28$ pp in the
concave region, pessimistic by $22$ pp in the convex tail) and the query average corrects both. The bound is \emph{safe
but loose} at low $\sigma_{\mathrm{loc}}$/small $L$, where $\bar h_\tau$ under-predicts measured recall by up to
$14$ pp. No retreat to the assumption-free envelope is needed; the only cost is that the guarantee is stated on
$\bar h_\tau$ (a query average) rather than on the mean density, and is conservative for spatially-correlated
predicates (binomial independence under-counts correlated in-predicate mass).

\paragraph{Bound-check at scale.} At $100$K, for the $c{=}1000$ merge the measured loss is $0.30$ pp, within the
batch-fraction cap $c/(N+c)=0.99$ pp ($3.3\times$ slack; at $18$K, $0.09$ pp within $5.26$ pp). The cap holds at
mid-scale, though the batch-fraction cap is weak (it is satisfied by any merge that does not corrupt a
constant fraction of the graph), so this is a sanity check, not a discriminating test ofS
Assumption~\ref{ass:permerge}. (The single-merge isolation at $c{=}10$K, $N{=}100$K independently reads $0.30$
pp, a coincidence of value from a different experiment.)

\subsection{Recall--latency feasibility}
\label{sec:eval-latency}
\begin{figure}[tbp]\centering\includegraphics[width=0.7\columnwidth]{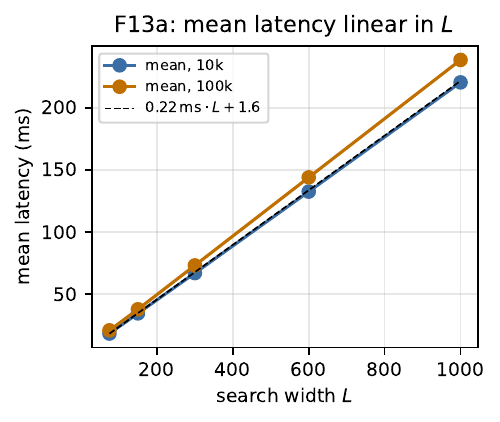}
\caption{Mean search latency is linear in the search width $L$ at both scales
($\mathrm{mean}(L)\approx0.22\,\mathrm{ms}\cdot L+1.6\,\mathrm{ms}$; mean disk IOs $\approx L$, one candidate
$\approx$ one IO), the IO-bound cost model on which the C1 admission test rests. Latencies are cache-empty
disk-index means; the figure establishes the linearity of the cost model, not absolute speed.}\label{fig:F13a}\end{figure}
The admission test of Contribution~C1 ($1.25\,k/\sigma_{\mathrm{loc}}\le L_{\max}(\ell_\tau)$) assumes search cost
is linear in $L$; we measure it. Mean search latency is linear in $L$ ($\mathrm{mean}(L)\approx0.22\,\mathrm{ms}
\cdot L+1.6\,\mathrm{ms}$, with mean disk IOs $\approx L$ exactly, since the search is IO-bound, one candidate
$\approx$ one IO; Fig.~\ref{fig:F13a}), so the cost model \textbf{holds} and $L_{\max}(\ell_\tau)\approx\ell_\tau/0.22\,\mathrm{ms}$, giving the
concrete feasibility condition
\begin{equation}
\begin{aligned}
  \sigma_{\mathrm{loc}}\ &\ge\ 1.25\,k\,/\,L_{\max}(\ell_\tau)\ \approx\ 1.25\,k\cdot 0.22\,\mathrm{ms}/\ell_\tau\\
  &\qquad(\text{e.g.\ }\ell_\tau=50\,\mathrm{ms}\Rightarrow\sigma_{\mathrm{loc}}\gtrsim0.055).
\end{aligned}
\end{equation}
This prices the boundary of \S\ref{sec:eval-strawmen}/\S\ref{sec:limitations} quantitatively: a
$\sigma_{\mathrm{loc}}=0.01$ tenant needs $L_{95}\approx1250$, i.e.\ $\approx275$ ms mean, so it cannot hold a
$95\%$ recall floor and an interactive latency SLO on this substrate, and the test reports this
\emph{before} deployment (Fig.~S2, supplementary material). We report the test on \textbf{mean} latency, with a $\rho{=}0.95$ floor and a candidate latency ladder
($\ell\in\{10,50,200\}$ ms). Since an interactive SLO is ultimately a \emph{tail} guarantee, a low-contention
$p_{99}$ capture would extend the test to the tail; the linear-cost and infeasibility conclusions here rest on the
mean.

\subsection{Multi-tenant SLA under churn (the payoff axis)}
\label{sec:eval-churn}
On an identical, genuinely-contended churn stream ($T{=}6$, tiers tight/medium/lax, ceilings $1000/3000/10^9$),
the C3 ceiling drives tight-tier SLA violations to \emph{zero} at equal staleness and equal drained recall
($82.34$ vs.\ $82.22$, i.e.\ enforcement costs nothing on terminal recall), where an otherwise-identical
scheduler-less baseline violates every tight query (Table~\ref{tab:churn}). The medium tier is uneventful in this
regime (zero violations on \emph{both} legs), so the ceiling protects the tight tier only. Two qualifications bound
the reading. The violation counted here is a per-tenant staleness-ceiling breach, staleness exceeding $S_\tau$,
a freshness bound distinct from the recall/latency SLO $(\rho_\tau,\ell_\tau)$. The baseline comparison is close
to definitional (C3 force-drains at $S_\tau$), so it establishes that the mechanism is implemented and effective; and the $100\%$-vs-$0\%$ split is a \emph{saturated} margin (tight queries arrive in a burst inside the
post-ceiling stale window), demonstrating the mechanism while the magnitude is specific to this $10$K,
three-checkpoint regime. The mechanism is not free in every regime; in an earlier configuration whose high-volume
tier was the \emph{lax} one, honoring tight ceilings deferred it and cost ${\approx}10$ pp of aggregate unfiltered
recall at intermediate checkpoints (final topology unaffected by merge-order invariance). The substantive
comparator is a per-tenant shard, against which ours is a rebuild-outage advantage: a shard is unqueryable for each
cold rebuild, growing with per-tenant size ($0.26$ s at $1.8$K/tenant to $\approx28$ s at $100$K/tenant), while
ours never rebuilds (no rebuild-induced serving outage; the index stays queryable through the background merge,
which does inflate the tail for $\approx c/d$ s but never takes the tenant offline). The shard's violation figures are \emph{modelled, not measured
head-to-head}: no query stream was run against a shard during rebuild. What is measured is the rebuild wall-clock;
the violation rate is the duty cycle $\min(1,r_{\mathrm{churn}}\!\cdot\!\mathrm{build}(n))$, an \emph{upper} bound
(it assumes uniform arrival and a query path that cannot dodge a rebuild, a churn-aware path would give $0$; we
lead with the upper bound because production paths do not know rebuild timing). At the measured $10$K scale the
shard's count is small (${\sim}3$/2000, $0.15$--$0.5\%$), so the direction is established
but the margin is thin and mid-scale sharding remains viable. And the magnitude at larger per-tenant sizes is
governed by the assumed churn rate. At the anchor's calibrated $r_{\mathrm{churn}}\approx10^{-2}$/s/tenant the model
reproduces the measured near-null, but sweeping $r_{\mathrm{churn}}$ from $1.7\times10^{-3}$ to $1.7\times10^{-2}$
moves the projected $100$K/tenant violation rate from $4.8\%$ to $48\%$, an order-of-magnitude spread set by the
assumed rate, not the build cost (measured to $100$K/tenant, and \emph{flattening} above ${\approx}60$K as the
parallel build saturates, so linear extrapolation would overstate the outage). The per-op wall-clock the shard wins
at small $n$ is \emph{unamortized}, since C3 folds ${\approx}6$ tenant-ops into one merge (narrowing the gap
${\approx}6\times$), and the shard's cost is a serving \emph{outage} while ours is background.

\begin{table*}[tbp]\centering\scriptsize
\caption{Tight-tier SLA violations under churn (per-tenant GT). Our rows and the baseline are measured; the shard row
is a duty-cycle \emph{upper bound} computed from measured rebuild wall-clocks, not a measured head-to-head.}
\label{tab:churn}
\begin{tabular}{lccc}
\toprule
system & tight viol & medium viol & note\\
\midrule
ours (C3 ceiling) & \textbf{0}/2000 & 0/3000 & force-drain, no rebuild\\
scheduler-less baseline & 2000/2000 & 0/3000 & no ceiling (medium not exercised)\\
per-tenant shard (UB) & ${\sim}3$/2000 & ${\sim}8$/3000 & rebuild-outage duty cycle @10K\\
\bottomrule
\end{tabular}
\end{table*}

\subsection{Baselines and Curator: the tradeoff on static data}
\label{sec:eval-strawmen}
On static low-per-tenant data ($T{=}6{\to}96$, fixed total $N$), per-tenant sharding is \emph{not}
navigability-limited (shards recall ${\approx}100$) but is \emph{memory}-limited, its summed footprint growing
linearly in tenant count and OOMing a $3$ GB box at $T{=}96$ (FreshDiskANN-shard $393{\to}6144$ MB, PipeANN-shard
$757{\to}11776$ MB), whereas our shared graph is flat. Under matched search cost our post-filter recall degrades
with tenant count (the $\sigma_{\mathrm{loc}}$ effect, \S\ref{sec:eval-h}). Our Curator comparison uses the authors' published artifact at a pinned commit, so it
measures genuine per-tenant navigability rather than a post-filter over a label-blind graph. It recalls
$\approx100$ flat at $\sim100$ MB in-memory and thus beats us on \emph{static} recall; our contribution is instead on streaming, per-tenant SLA scheduling, the bound and
disk-resident scale, axes the evaluated static, in-memory Curator artifact does not target. The outcome on static data is a
\emph{tradeoff}, not a clean win for either side; the payoff axis is churn (\S\ref{sec:eval-churn}) and scale
(\S\ref{sec:eval-scale}).

\subsection{Scale: bounded and correct at mid-scale; the inversion beyond}
\label{sec:eval-scale}
At $100$K our prototype runs \emph{bounded} (steady RSS $\approx1.06$ GB single-tenant, $\approx1.67$ GB
six-tenant) and \emph{correct} (recall@10 $99.70$, on-curve), with the SLA win intact, closing the one
at-scale assumption the projection could not vouch for (Table~\ref{tab:scale}). Because in-memory footprints
are $\propto N$ while a disk-resident index keeps the graph and full-precision vectors on disk, the memory picture
should \emph{invert} at scale, since an in-memory Curator at its measured $5.6$ KB/vector exceeds $16$ GB by
${\approx}2.9$M points and needs ${\approx}56$ GB at $10$M, whereas only the PQ-compressed vectors and tags
($32$--$128$ B/vector) grow with $N$ on our side, putting us at ${\approx}1$--$2$ GB even at $10$M.

\emph{The scaling claim rests on a decomposition, not on the two headline RSS points, which conflate several
quantities.} An RSS breakdown at fixed tenant and thread count ($T{=}6$, $32$ threads) on the matched $18$K/$100$K
pair separates three marginals. The \emph{serving} graph-residency marginal is \textbf{0.115 kB/point} (PQ codes
and tags resident in RAM, the graph on disk), matching the PQ-code size architecturally ($\approx102$ B/point at
$100$K); serving memory is therefore $N$-independent up to that marginal and the bounded-memory claim holds
\emph{for serving}. The \emph{merge-peak transient} is \textbf{8.3 kB/point}, incurred while the StreamingMerger
rewrites the $O(N)$ graph in RAM under $32$ build threads and released afterwards (proportional to $N$, reducible
by thread and batch configuration). Between the two sits the \emph{whole-process first stable RSS}, which grows at
$\approx2.6$ kB/point on the matched pair, above the serving marginal because it includes configured allocations
(per-thread search buffers, staging sized as $\#\text{partitions}\times2\,\mathrm{Merge\_Size}$) that the supplementary RSS decomposition (Fig.~S3) does not
fully attribute to $N$; isolating that residual is left open. This decomposition supersedes an earlier reading
that took the raw two-point $18$K$\to$$100$K RSS slope ($\approx10$ kB/point) at face value and extrapolated it to
$\approx11$ GB at $1$M. That slope is the mixed whole-process quantity, not a serving cost, which is why the
$1$M--$10$M rows in Table~\ref{tab:scale} are quoted as residency-model \emph{projections} on the $0.115$ kB/point
serving term rather than as extrapolations of the raw curve. The $0.115$ kB/point marginal is a two-point
measurement whose agreement with the PQ-code size makes it credible.

\begin{table*}[tbp]\centering\scriptsize
\caption{Memory vs.\ $N$ on the matched $T{=}6$ pair, on \emph{consistent} conventions:
\emph{serving} = whole-process first stable RSS (its per-vector PQ+tags component is $0.115$ kB/point);
\emph{merge peak} = process peak RSS (a released transient, $8.3$ kB/point). Curator is in-memory
($\propto N$). $1$M/$10$M are residency-model projections; serving$^\ddagger$ is quoted on the PQ+tags term, the
$1$M merge peak is unrun. (The single-tenant $100$K point, $\approx1.06$ GB, is
Fig.~\ref{fig:F11F12}a; it is not a $T{=}6$ pair member and is omitted here to keep conventions matched.)}
\label{tab:scale}
\begin{tabular}{lcccc}
\toprule
$N$ & ours serving & ours merge-peak & Curator (in-mem) & status\\
\midrule
18{,}000 ($T{=}6$)  & 0.77 GB & 1.04 GB & 0.10 GB & measured\\
100{,}000 ($T{=}6$) & 0.98 GB & 1.72 GB & --- & measured\\
1{,}000{,}000  & $\approx1.0$ GB$^\ddagger$ & $\approx9$ GB & 5.6 GB & projected\\
10{,}000{,}000 & $\approx2$ GB$^\ddagger$ & --- & 56 GB (OOM) & projected\\
\bottomrule
\end{tabular}\\[2pt]
{\raggedright\footnotesize $^\ddagger$ \textbf{Convention (measured vs projected serving differ and must be read
together).} The measured $18$K/$100$K serving cells are \emph{whole-process} first stable RSS (fixed baseline
$\approx0.9$ GB + configured allocations + the PQ+tags residency); across the matched pair this rises at
$\approx2.6$ kB/point. The projected $1$M/$10$M serving cells grow the \emph{only $\propto N$ term}, the PQ+tags
residency at $0.115$ kB/point (matching the PQ-code size architecturally, $\approx102$ B/point at $100$K), holding
the bounded remainder fixed, justified because the RSS decomposition (Fig.~S3) attributes the $2.6$ kB/point
excess to the C2 cache (bounded by high-water, \emph{not} $\propto N$) and per-thread buffers (fixed), not to
$N$-growth. Hence the projection $\approx0.9$ GB baseline $+\,0.115$ kB/point$\,\cdot N$. A naive linear
extrapolation of the $2.6$ kB/point two-point whole-process slope gives the pessimistic $\approx3.4$ GB at $1$M;
we report the architectural term as primary because the excess is measured-bounded, and both stay far under the
in-memory $\propto N$ baseline. The two-point marginal is not yet firmed by a third matched point.\par}
\end{table*}
\begin{figure}[tbp]\centering
\includegraphics[width=0.7\columnwidth]{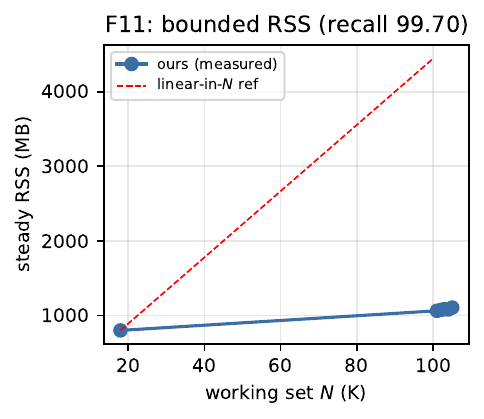}\\[6pt]
\includegraphics[width=0.73\columnwidth]{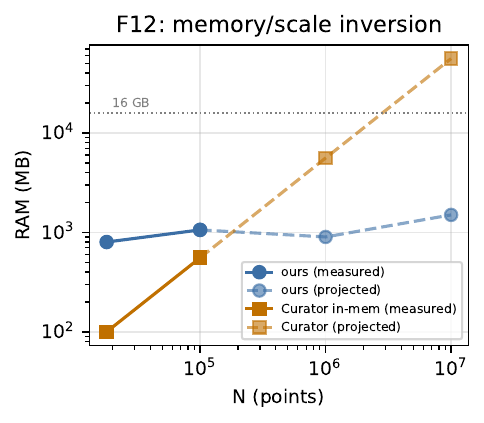}
\caption{(a, top) Does-it-run: the prototype is bounded and correct at $100$K ($\approx1.06$ GB, recall $99.70$ on-curve). (b, bottom) Memory/scale inversion (log-log): a disk-resident index keeps graph and full vectors off-RAM so only PQ codes grow with $N$, while an in-memory index scales $\propto N$; measured to $100$K, \emph{modelled} at $1$M--$10$M, with the like-for-like two-point slope shown as the pessimistic alternative.}\label{fig:F11F12}\end{figure}

\subsection{External comparisons}
\label{sec:eval-external}
Preliminary cross-system comparisons (single-tenant streaming vs.\ PipeANN, filtered recall vs.\
Filtered-DiskANN, and tail latency) are reported in the supplementary material. They are preliminary (the PipeANN gap is on unmatched parameters, a param-matched head-to-head ongoing) and are not central to
our claims.

\section{Related work}
\label{sec:related}

\paragraph{Filtered-static graph ANN.} Filtered-DiskANN~\cite{filtereddiskann} and ACORN~\cite{acorn} encode
predicates into graph construction; RNSG~\cite{rnsg} and iRangeGraph~\cite{irangegraph} target range predicates;
STABLE~\cite{stable} fuses metric and attribute signals into one joint-semantic graph. All build a \emph{single
static}, single-tenant index over the joint space; we add streaming inserts/deletes/tenancy-migration, multi-tenant
isolation, and an analytical bound. These designs secure same-label reachability at build time; our repair must
preserve it under churn, which shapes the scheduler's ordering.

\paragraph{Streaming / dynamic graph ANN.} FreshDiskANN~\cite{freshdiskann} (two-pass StreamingMerge, our
substrate), SPFresh~\cite{spfresh}, IP-DiskANN~\cite{ipdiskann}, CleANN~\cite{cleann}, Wolverine~\cite{wolverine},
and Mycelium-Index~\cite{myceliumindex} operate on unfiltered, single-tenant data with at best empirical recall
stability. PipeANN~\cite{pipeann} (OSDI'25) and OdinANN~\cite{odinann} (FAST'26) are the current single-tenant SOTA,
their pipelined-SSD search and direct-insert (no-merge) design sidestepping the scheduling problem we solve rather
than solving it; neither has a tenant concept or a recall bound, so we cede the single-tenant axis and compare
against a per-tenant \emph{sharding} of each. LSM-VEC~\cite{lsmvec} organizes a vector index as an LSM tree, and
UBIS~\cite{updatablebalanced} is the closest streaming competitor, scheduling concurrent updates to keep recall
stable under high update rates but reporting that empirically with no bound. None provides filtering $\times$
multi-tenancy $\times$ a bound jointly.

\paragraph{Multi-tenant ANN.} Curator~\cite{curator} is the closest prior art, a multi-tenant graph ANN with
per-tenant navigability (clustering sub-trees) that is static and in-memory, with no per-tenant SLA scheduling and
no bound; we build it and compare (conceding static recall, leading on streaming/SLA/bound/disk-scale). Pinecone
Metadata Filtering~\cite{pineconefilter} provides filter + multi-tenant + streaming on an IVF/serverless substrate
with no analytical bound; GaussDB-Vector~\cite{gaussdbvector} is a production filtered vector database with
real-time inserts/deletes but no per-tenant isolation, SLA scheduling, or bound.

\paragraph{Caches over vector indexes.} MicroNN~\cite{micronn}, QVCache~\cite{qvcache}, and
LiveVectorLake~\cite{livevectorlake} are read-side, or filter-/tenant-blind, or a different layer; our cache is
jointly filter and tenant-keyed, on the query path, and is the structure the theorem binds.

\paragraph{Theoretical bounds.} ~\cite{quantstream} bounds \emph{quantization} under streaming updates, not a
graph's recall; Random-Walk Dynamic Graph ANN~\cite{randomwalkann} bounds hitting-time topology preservation, not
recall@$k$ under an SLA; Quake~\cite{quake} estimates recall for \emph{query-time} parameters, not background work.
Distribution-Aware HNSW~\cite{distawarehnsw} informs our predicted-recall-impact scheduler term. Ours is, to our
knowledge, the first per-tenant recall-loss framework for the joint filtered, multi-tenant, streaming graph-ANN
setting, a decomposition unconditional in three of its four terms, with the staleness term instantiated from a
degradation law calibrated on this substrate.

\paragraph{Benchmarks.} We compose existing benchmarks rather than propose one, drawing on the
Big-ANN'23~\cite{bigann23} streaming track with filtered slices alongside CANDY~\cite{candy},
CANDOR-Bench~\cite{candorbench}, and the Filtered-ANN Unified Benchmark~\cite{filteredannbench}. The substrate
builds on Vamana/DiskANN~\cite{diskann}, HNSW~\cite{hnsw}, NSG~\cite{nsg}, and SPANN~\cite{spann}.

\section{Discussion}
\label{sec:limitations}

\paragraph{The staleness half of the bound is semi-empirical.} Its mechanism and functional form are proved and
validated on two independent axes, while the degradation exponents ($\theta\approx1.46$, $b\approx0.5$) are
\emph{calibrated} rather than derived from first principles. Three properties bound what it establishes, all
quantified in \S\ref{sec:toy} and \S\ref{sec:eval-accum}. The fit is valid over $0.05\le c/N\le0.45$, with the
smallest batch sizes falling under the $0.02$ pp reporting floor. The exponent CI ($a_{\mathrm{recall}}{=}1.28$,
95\% CI $[1.03,1.52]$) supports $a-1>0$ but not its magnitude, so the design rests on the \emph{sign} $a\ge1$
rather than the value. And the horizon assembly is an upper bound loose by $2.3$--$2.7\times$, since $\eta<1$. The
measured monotonicity of $L^*(c)$ rules out the one qualitative failure mode, leaving the deployed-regime magnitude
weakly constrained but not the direction.

\paragraph{The filtered term is a loss bound only under homogeneous local density.} Lemma~\ref{lem:frontier}(ii)
is parameter-free ($\beta{=}1$ is forced by the engine contract) and predicts the recovery constant rather than
fitting it. For spatially-clustered predicates the two-sided bound is validated by evaluating the query average
$\bar h_\tau$ on the archived per-query $\sigma_{\mathrm{loc}}$ (\S\ref{sec:eval-h}), which absorbs the up-to-$26$ pp
mean-substitution gap and holds as a \emph{conservative} lower bound on delivered recall (measured
$\ge\bar h_\tau$ in $14$ of $15$ clustered cells), loose by up to $14$ pp at low $\sigma_{\mathrm{loc}}$. Where
per-query densities are unavailable, the assumption-free envelope Lemma~\ref{lem:frontier}(i) constrains the
required search width but gives no per-tenant recall guarantee.

\paragraph{Filtered recall degrades with $\sigma_{\mathrm{loc}}$ by design.} On the label-blind graph, recall falls
as local density falls (Lemma~\ref{lem:frontier}), the theorem's content, recoverable at
$L=\Theta(k/\sigma_{\mathrm{loc}})$ but conditional on the tenant's latency SLO via the feasibility test
(\S\ref{sec:eval-h}). A tenant at $\hat\sigma_{\mathrm{loc}}\approx0.01$ needs $L\approx1250$ and cannot hold a
$95\%$ recall floor and an interactive SLO at once, for which a per-tenant-navigable index is the right structure.
This does not refute the multi-tenant thesis; the binding variable is predicate \emph{diffuseness}, not tenant
count ($\hat\sigma_{\mathrm{loc}}$ exceeds $1/T$ by $3$--$5\times$ for spatially-coherent tenants, \S\ref{sec:eval-h}).
The feasibility test is reported on mean rather than tail latency.

\paragraph{Scale and measurement scope.} The SLA \emph{mechanism} is measured (ceiling enforcement drives
tight-tier violations to zero under churn where a scheduler-less baseline cannot) and the shard rebuild-outage is
measured to $100$K/tenant, but the shard violation \emph{rate} is a duty-cycle model and the $1$M--$10$M memory
inversion is a projection from measured per-vector RSS. The topology-$\Delta$ is a same-method \emph{relative}
measurement with limited power to exclude sub-pp
differences, and our absolute $p_{99}$ trails PipeANN's by roughly an order of magnitude at the configurations
measured; external comparisons are preliminary (the PipeANN recall gap is on unmatched parameters; ACORN did not
build within our memory ceiling). Results are on a single dataset family (SIFT and a filtered YFCC slice) and a
single storage backend, so the measured constants may not transfer to other datasets or storage stacks without
recalibration. Finally, the analysis presupposes a merge-based (out-of-place) substrate; direct-insert indexes have
no $c$/$d$/$\tau$ referent and lie outside our scope.

\section{Conclusion}
\label{sec:conclusion}

We studied the intersection of graph-substrate, filtered, multi-tenant, streaming ANN with an analytical per-tenant
recall@$k$ bound, which prior work has not jointly addressed. The organizing result is anti-amortization, and we
observe it directly: over a fixed $60$K-insert horizon and a $20\times$ range of merge sizes, terminal recall falls
monotonically as merges grow. On this substrate, batching is therefore not a free recall-amortization lever;
degradation can be re-prioritized but not amortized away. The per-merge law behind it is super-linear in $c$ across
the evaluated $c/N$ range, with an exponent estimated from that range rather than derived. That fact explains the
design: a filter and tenant-aware staging cache that absorbs within-cycle staleness on the query path (C2), and a
select-not-resize, SLA-aware scheduler that re-orders repair under a per-tenant max-staleness ceiling (C3), both
binding to a bound we generalize to the filtered (local-density $\sigma_{\mathrm{loc}}$), multi-tenant
(cadence-bounded), deletes/migration setting (C1). The filtered half of that bound is a parameter-free
frontier-mass law (expected in-predicate frontier mass truncated at $k$) that predicts the measured recovery
scale $L_{95}=1.25\,k/\sigma_{\mathrm{loc}}$ and holds as a loss bound under homogeneous local density. On a
FreshDiskANN substrate we measure that the scheduler's re-ordering leaves terminal recall unchanged, that
the per-tenant staleness ceiling eliminates tight-tier violations under churn where a scheduler-less baseline
cannot, and that the design stays memory-bounded and recall-correct at mid-scale. PipeANN and OdinANN remain stronger on single-tenant recall and tail latency, and
Curator on static filtered recall; our contribution targets the filtered, multi-tenant, streaming axes and the
per-tenant freshness guarantee.


\bibliographystyle{IEEEtran}
\bibliography{refs}

\end{document}


\title{Supplementary Material for\\``Recall-Bounded Staleness for Filtered Multi-Tenant ANN Graph Indexes under Concurrent Updates''}

\author{Namrata~Das, Kailas~Devadkar, Anand~Godbole\\
Sardar Patel Institute of Technology, Mumbai, India}

\markboth{Das \MakeLowercase{\textit{et al.}}: Recall-Bounded Staleness (Supplementary Material)}%
{Das \MakeLowercase{\textit{et al.}}: Recall-Bounded Staleness (Supplementary Material)}

\maketitle

\noindent This document supplements the main manuscript. It contains the full proofs
(Appendix~\ref{app:proofs}), three supporting tables, external comparisons, and three supporting figures.
Numbering of results, sections, and tables referenced below (e.g.\ Theorem~3, Lemma~7,
\S\,V-C) is that of the main paper.

\section{Proofs}
\label{app:proofs}

\subsection{Proof of the toy bound (Theorem~3)}
\label{app:toy}
The load-bearing object is the per-merge degradation $g(c,N)$ (Assumption~2); the horizon bound
then assembles $M$ merges and concentrates. We couple the merged graph $G^{(c)}_{\mathrm{merge}}$ against the
from-scratch $\alpha$-RNG-optimal graph on the same final point set.

\emph{(S1--S3) Topology divergence.} By Definition~1 the $c$ batch points are mutually invisible, so
every intra-batch shortcut a from-scratch build would create is missing. The count of such missing edges is
\begin{equation}\label{eq:emiss}
E_{\mathrm{miss}}=\Theta\!\big(\min(c^2/N,\,Rc)\big),
\end{equation}
the $R$-cap on out-degree squeezing the naive $c^2/N$ below the linear ceiling $Rc$
(measured $R$-cap fill $15\%\!\to\!50\%$ over the $c$-sweep; fixed-$N$ exponent $\theta{=}1.46$, 95\% CI
$[1.33,1.59]$ over $n{=}5$ values of $c$). The fill measurement fixes the form, placing the exponent
strictly between the unconstrained $c^2$ and the saturated $Rc$ and making the fitted slope range-dependent
($1.34$ over $c\in[5,20]$K, $1.46$ over $c\in[5,80]$K) rather than a universal constant.
The hub re-prune (M2) contributes a subdominant, budget-soft $V_0$ floor $g_0(L)$.

\emph{(S4) Divergence $\to$ recall.} The topology divergence lifts to recall@$k$ loss near-linearly
($\psi\approx1$) below the batch-fraction ceiling $c/(N+c)$. Three independent measurements agree on the exponent.
Edge counting gives $\theta{=}1.46$; a three-replicate streaming recall sweep gives $\gamma{=}1.46\pm0.14$; a
$\mu$-matched three-base recall fit gives $a_{\mathrm{recall}}{=}1.45$ with $b_{\mathrm{recall}}{=}0.64$ against
the topology $b{=}0.50\pm0.04$. The $\pm10$--$15\%$ spread across methods fixes $\psi\approx1$ as near-linear lift
with no amplifying power. Matching $\mu=kc/N$ across bases aligns the transfer, leaving a mild $0.50$ vs.\ $0.64$
residual between the topology and recall exponents.

\emph{(S5) Horizon assembly.} Summing $g(c,N_m)$
over the\\ $M=\lceil\lambda_{\mathrm{ins}}T/c\rceil$ merges with growing $N_m$ gives
\begin{equation}\label{eq:lstar}
L^*(c,T)=\Theta\!\big(\lambda_{\mathrm{ins}}T\,c^{a-1}N_0^{-b}\big),
\end{equation}
which increases in $c$ because $a-1>0$, the anti-amortization direction. A direct partition-counting argument,
with total missing intra-batch edges $\Theta(\lambda_{\mathrm{ins}}Tkc/N)\propto c$, re-derives this direction
independently of the coupling. The merges are combined by a union bound over miss events, giving
\begin{equation}\label{eq:lstar-ub}
L^*\le\sum_m g(c,N_m),
\end{equation}
the upper bound Theorem~8 uses. Reconcile only unions buffered reverse edges and re-prunes, so it leaves earlier
missing intra-batch edges in place and the per-merge terms accumulate. Loss components share capacity across
merges, which makes the additive sum~\eqref{eq:lstar-ub} conservative. Over twelve merges (\S\,V-B) the realized
cumulative loss is $0.37$--$0.44$ of the additive prediction, and the $c$-independent $V_0$ floor decays with merge
index. This saturation of the $c$-independent floor removes the $M g_0\propto1/c$ term, so it eliminates rather
than creates an amortization regime.

\emph{(S6) Concentration.} The degree cap gives bounded per-insert influence $O(R^2)$. McDiarmid on the divergence
and Bernstein on the sparse, bounded ($X_q\in[0,1]$) query-recall average, union-bounded over the $M$ merge
checkpoints, yield
\begin{equation}\label{eq:conc}
L^*\le\mathbb{E}[L^*]+O\!\big(R^2\sqrt{\lambda_{\mathrm{ins}}T\ln(M/\delta)}\big)=\mathbb{E}[L^*](1+o(1))
\end{equation}
with probability $\ge1-\delta$. The exponents $\theta$ and $b$ are calibrated on this substrate (Remark~9), and the
direction and functional form follow from the argument above and its measurements.

\subsection{Proof of the frontier-mass lemma (Lemma~7)}
\label{app:frontier}
\emph{(S0) The recall identity.} By Assumption~4 the post-filter receives
a distance-ordered list of $\beta L$ candidates with in-predicate count $X_q$. Suppose first the list is the
\emph{true} metric top-$\beta L$. Its in-predicate members are then exactly the $X_q$ nearest in-predicate points
of the whole dataset, since any nearer in-predicate point would itself lie in the top-$\beta L$. The
per-label ground truth is the $k$ nearest in-predicate points, so the retained top-$\min(X_q,k)$ of the filtered
list is a prefix of the ground-truth list, giving recall $=\min(X_q,k)/k$ as an exact equality rather than a
bound. This is the quantity the measurement of \S\,V-C computes, taking the post-filter top-$k$ and scoring
against per-label brute-force ground truth. For an approximate search the list is
not the true top-$\beta L$; write $\delta_{\mathrm{ann}}(L)$ for the expected fraction of a query's in-predicate
ground-truth points that lie in the true top-$\beta L$ yet are absent from the returned list. Each such point costs
at most $1/k$ of recall, so
\begin{equation}\label{eq:recall-lb}
\mathbb{E}[\mathrm{recall}]\ge\mathbb{E}[\min(X_q,k)]/k-\delta_{\mathrm{ann}}(L).
\end{equation}
The argument~\eqref{eq:recall-lb} bounds the \emph{shortfall} and says nothing about recall exceeding the
frontier-mass value, so the inequality is one-sided. One-sided suffices because Theorem~8 adds
$\delta_{\mathrm{ann}}(L)$ to an upper bound on loss, and $\delta_{\mathrm{ann}}$ is measurable against the
archived brute-force neighbour lists. The object is a truncated mean rather than a success probability, the
distinction that separates parts (i)--(iv) below.

\emph{(i) Envelope.} The map $u\mapsto\min(u,k)$ is concave, so by Jensen and $\mathbb{E}X=\beta L\sigma_{\mathrm{loc}}$,
\begin{equation}\label{eq:env}
\mathbb{E}[\min(X,k)]/k\le\min(1,\beta L\sigma_{\mathrm{loc}}/k).
\end{equation}
No assumption on the law of $X$ beyond its mean is used, so~\eqref{eq:env} holds for arbitrarily correlated
predicates. Recall $\to1$ therefore requires $\beta L\sigma_{\mathrm{loc}}\ge k$, i.e.\
$L=\Omega(k/\sigma_{\mathrm{loc}})$, and with (ii) the recovery scale is $\Theta(k/\sigma_{\mathrm{loc}})$.

\emph{(ii) Exact form under homogeneity.} Under Assumption~6,
$X\sim\mathrm{Binom}(\beta L,\sigma_{\mathrm{loc}})$. For a non-negative integer variable
$\mathbb{E}[\min(X,k)]=\sum_{i=1}^{k}\Pr[X\ge i]$, hence
\begin{equation}\label{eq:h}
h=\frac1k\sum_{i=1}^{k}\Pr[\mathrm{Binom}(\beta L,\sigma_{\mathrm{loc}})\ge i].
\end{equation}
Every summand is $\ge\Pr[X\ge k]$, so $h\ge h_{\mathrm{tail}}=\Pr[X\ge k]$. The single-term tail is a valid
lower bound on recall but only the smallest of $k$ terms, and is loose by orders of magnitude at
$\beta L\sigma_{\mathrm{loc}}\ll k$ (\S\,V-C, Table~\ref{tab:hfit}). Combining with~\eqref{eq:env},
\begin{equation}\label{eq:sandwich}
h_{\mathrm{tail}}\le h\le\min(1,\beta L\sigma_{\mathrm{loc}}/k).
\end{equation}
At the limits $h\to1$ when $\beta L\sigma_{\mathrm{loc}}\gg k$ (all $k$ tail terms saturate) and
$h\to\beta L\sigma_{\mathrm{loc}}/k$ when $\beta L\sigma_{\mathrm{loc}}\ll k$ (the truncation is inactive,
$\mathbb{E}[\min(X,k)]\to\mathbb{E}X$), the latter being the unit-slope regime that calibrates $\beta{=}1$.

\emph{(iii) Query average.} A tenant's recall is \\
$\bar h_\tau=\mathbb{E}_{q\mid\tau}[h(\sigma_{\mathrm{loc}}(q),L,k)]$, from (ii) applied per query and averaged. Put
$\gamma(p)=\mathbb{E}[\min(\mathrm{Binom}(n,p),k)]$ with $n=\beta L$. Differentiating the binomial expectation
twice in $p$ gives $\gamma''(p)=n(n-1)\,\mathbb{E}[\Delta^2\phi(\mathrm{Binom}(n-2,p))]$, where $\phi(j)=\min(j,k)$
and $\Delta^2\phi(j)=\phi(j+2)-2\phi(j+1)+\phi(j)\le0$ by concavity of $\phi$, so $\gamma$ is concave on $[0,1]$. By
Jensen,
\begin{equation}\label{eq:qavg}
\bar h_\tau=\mathbb{E}_q[\gamma(\sigma_{\mathrm{loc}}(q))]/k\le\gamma(\mathbb{E}_q[\sigma_{\mathrm{loc}}(q)])/k
=h(\bar\sigma^{(\tau)}_{\mathrm{loc}},L,k).
\end{equation}
Substituting a tenant's mean local density therefore over-predicts its recall, and the gap is governed by the
dispersion of $\sigma_{\mathrm{loc}}$ across the tenant's queries. The gap is one-signed and vanishes wherever
$\gamma$ is affine, in the unsaturated regime $\gamma(p)\to np$ (linear) and the saturated regime $\gamma\to k$
(constant), so it peaks in the knee, matching \S\,V-C. Because $\bar h_\tau$ is computable from the per-query
$\sigma_{\mathrm{loc}}$ the sweep already produces, it is a quantitative prediction (\S\,VII).

\emph{(iv) Ordering.} Every summand of (ii) is $\ge\Pr[X\ge k]$ and, by~\eqref{eq:env}, $\gamma(p)/k\le\min(1,np/k)$;
combining with (iii),
\begin{equation}\label{eq:order}
h_{\mathrm{tail}}\le\bar h_\tau\le h(\bar\sigma_{\mathrm{loc}},L,k)\le\min(1,\beta L\bar\sigma_{\mathrm{loc}}/k).
\end{equation}
The tail $h_{\mathrm{tail}}$ lower-bounds recall unconditionally and the envelope upper-bounds it
unconditionally.\hfill$\square$

\subsection{Multi-tenant extension (Theorem~8)}
\label{app:mt}
By Assumption~5 the staleness/degradation term is filter- and tenant-independent, so it applies
per tenant with $N_\tau^{\mathrm{stale}}$ in place of the global insert count; the scheduler's max-staleness ceiling
caps $N_\tau^{\mathrm{stale}}\le\lambda^{(\tau)}_{\mathrm{ins}}(S_\tau+W)$ for a trigger window $W$, a scheduler
invariant and the ceiling-test granularity of \S\,IV-C.
The filtered term is the query-averaged Lemma~7(ii)--(iii) on the fixed graph. Additivity follows from a
union bound. Fix a ground-truth item $y$ for query $q$; let $A$ be the
event that $y$ is a pending mutation neither integrated into $G_t$ nor served by the cache, and $B$ the event that
$y$ is absent from the post-filtered frontier of the converged graph. The item is missed only on $A\cup B$, so
\begin{equation}\label{eq:mt-union}
\mathbb{E}[1-R_q]\le\Pr[A]+\Pr[B],
\end{equation}
and averaging over $y$, $q$ and taking the supremum over $t$ gives the displayed inequality, with slack
$\Pr[A\cap B]$ from items that are both stale and out-of-frontier. Parameter disjointness makes the two terms
separately \emph{estimable}. Deletes and migration add a $\lambda_{\mathrm{del}}$-dependent term $g_0^{\mathrm{del}}$
to the M2 hub-perturbation floor. The high-probability form extends by a union bound over $|\mathcal{T}|$ tenants,
since the $R$-cap bounded influence survives the union. Evaluating a tenant's filtered term at its mean local
density rather than the query average $\bar h_\tau$ replaces the bound with the estimate of Lemma~7(iii); the
assumption-free fallback is Lemma~7(i), giving a necessary-$L$ condition per tenant.\hfill$\square$

\section{Supplementary tables}
\label{sec:supp-tabs}
Three tables back the proofs and the evaluation. Table~\ref{tab:permerge} gives the per-merge degradation
$g(c,N_0)$ that underlies the missing-edge count~\eqref{eq:emiss} and the horizon sum~\eqref{eq:lstar}.
Table~\ref{tab:accum} tracks the horizon assembly that makes the additive bound~\eqref{eq:lstar-ub} conservative,
reporting realized cumulative loss against the additive prediction. Table~\ref{tab:hfit} compares the frontier-mass
model $h$ of~\eqref{eq:h} against measured recall on the collapsed curve, alongside the single-term tail
$h_{\mathrm{tail}}$ and the envelope~\eqref{eq:env}.

\begin{table}[htbp]\centering\scriptsize
\caption{Per-merge degradation at $N_0{=}100$K, $L{=}75$, recall@10, $10$K queries, cache-empty Tier-0b. Rows below
the $0.02$ pp reporting floor are excluded from the fit.}
\label{tab:permerge}
\begin{tabular}{rrrrrr}
\toprule
$c$ & $c/N_0$ & $R_{\mathrm{scratch}}$ & $R_{\mathrm{merge}}$ & $g$ (pp) & cap $c/(N_0{+}c)$ \\
\midrule
1{,}000  & 0.01 & 99.71 & 99.722 & $\approx0.00$\rlap{$^\dagger$} & 0.0099 \\
2{,}000  & 0.02 & 99.73 & 99.711 & 0.019\rlap{$^\dagger$}         & 0.0196 \\
5{,}000  & 0.05 & 99.72 & 99.633 & 0.087 & 0.0476 \\
10{,}000 & 0.10 & 99.73 & 99.534 & 0.196 & 0.0909 \\
20{,}000 & 0.20 & 99.70 & 99.340 & 0.360 & 0.1667 \\
40{,}000 & 0.40 & 99.62 & 98.500 & 1.120 & 0.2857 \\
80{,}000 & 0.80 & 99.53 & 96.494 & 3.036 & 0.4444 \\
\bottomrule
\end{tabular}\\[2pt]
{\raggedright\scriptsize $^\dagger$ at or below the $0.02$ pp floor set by 2-decimal reporting of
$R_{\mathrm{scratch}}$; excluded from all fits. This floor coincides with the $c/N\le0.02$ regime, so the
fitted per-merge law is quoted for $c/N>0.02$.\par}
\end{table}

\begin{table}[htbp]\centering\footnotesize
\caption{Horizon assembly at $c{=}5000$, $N_0{=}100$K. Measured cumulative Tier-0b loss vs.\ the additive
prediction $M\!\cdot\!g$.}
\label{tab:accum}
\begin{tabular}{rrrrr}
\toprule
$m$ & live $N$ & cum.\ loss (pp) & $M\!\cdot\!g$ (pp) & ratio $\eta$ \\
\midrule
1  & 105{,}000 & 0.000 & 0.087 & --- \\
4  & 120{,}000 & 0.188 & 0.348 & 0.54 \\
8  & 140{,}000 & 0.292 & 0.696 & 0.42 \\
12 & 160{,}000 & 0.384 & 1.044 & 0.37 \\
\bottomrule
\end{tabular}
\end{table}

\begin{table}[htbp]\centering\scriptsize\setlength{\tabcolsep}{3.5pt}
\caption{Model comparison on the collapsed curve. $\mu=\sigma_{\mathrm{loc}}L$, $x=\mu/k$, $k{=}10$; all columns in
recall-\%. $h$ is Lemma~7(ii) with $\beta{=}1$ (no fitted constant); $h_{\mathrm{tail}}$ is the
single-term tail with the previously-calibrated $\beta{=}1.66$; envelope is Lemma~7(i).
The $h$ column is computed from the 5-decimal $\hat\sigma_{\mathrm{loc}}$ using a normal
approximation to $\mathbb{E}[(k-X)^+]$ rather than the exact binomial; the approximation is good to
$\approx\pm0.2$ pp, so per-cell residuals below that magnitude are within numerical precision and the ${\le}1.2$ pp
headline is quoted at that precision.}
\label{tab:hfit}
\begin{tabular}{lcccccc}
\toprule
$(\hat\sigma_{\mathrm{loc}},L)$ & $\mu$ & $x$ & meas. & $h$ & resid & $h_{\mathrm{tail}}$\\
\midrule
\multicolumn{7}{l}{\emph{diffuse} (ladder; $\hat\sigma_{\mathrm{loc}}$ measured, $\approx1/T$)}\\
$(0.01033,75)$   &  0.775 & 0.078 &  7.75 &  7.75 & $-0.00$ & ${<}0.01$\\
$(0.02077,75)$   &  1.558 & 0.156 & 15.63 & 15.58 & $-0.05$ & ${<}0.05$\\
$(0.01033,300)$  &  3.099 & 0.310 & 30.98 & 30.99 & $+0.01$ & 2.6\\
$(0.04168,75)$   &  3.126 & 0.313 & 31.36 & 31.25 & $-0.11$ & 3.8\\
$(0.01033,600)$  &  6.198 & 0.620 & 61.41 & 61.31 & $-0.10$ & 55.8\\
$(0.02077,300)$  &  6.231 & 0.623 & 61.03 & 61.63 & $+0.60$ & 58.9\\
$(0.08383,75)$   &  6.287 & 0.629 & 61.87 & 62.24 & $+0.38$ & 62.2\\
$(0.01033,1000)$ & 10.330 & 1.033 & 89.99 & 88.83 & $-1.16$ & 96.0\\
$(0.04168,300)$  & 12.504 & 1.250 & 95.29 & 95.25 & $-0.04$ & 99.7\\
$(0.16750,75)$   & 12.563 & 1.256 & 95.42 & 96.07 & $+0.66$ & 99.7\\
$(0.02077,1000)$ & 20.770 & 2.077 & 99.94 & 99.87 & $-0.07$ & 100.0\\
\midrule
\multicolumn{7}{l}{\emph{clustered} ($h$ at the mean $\hat\sigma_{\mathrm{loc}}$; all residuals one-signed)}\\
$(0.04702,75)$   &  3.527 & 0.353 & 29.18 & 35.26 & $+6.1$  & 5.6\\
$(0.04702,150)$  &  7.053 & 0.705 & 47.17 & 68.88 & $+21.7$ & 71.4\\
$(0.04702,300)$  & 14.106 & 1.411 & 71.61 & 97.58 & $+26.0$ & 100.0\\
$(0.04702,600)$  & 28.212 & 2.821 & 92.80 & 100.0 & $+7.2$  & 100.0\\
$(0.16779,75)$   & 12.584 & 1.258 & 73.55 & 96.09 & $+22.5$ & 99.7\\
$(0.16779,150)$  & 25.169 & 2.517 & 92.30 & 100.0 & $+7.7$  & 100.0\\
$(0.51053,75)$   & 38.290 & 3.829 & 97.78 & 100.0 & $+2.2$  & 100.0\\
\bottomrule
\end{tabular}
\end{table}

\section{External comparisons}
\label{sec:supp-external}
On single-tenant streaming, recall@10 is $99.29/99.15/99.02$ over res $20/40/60$K against PipeANN's $99.94$, at
unmatched parameters (ours $R{=}32/L{=}75$/beam$4$ vs.\ PipeANN $R{=}64/L{=}100$/beam$32$). Tail latency is
$p_{99}\approx45$ ms against PipeANN's $3.1$--$6.1$ ms, the regime the pipelined-SSD search targets and the
beam-$4$ \texttt{io\_uring} path does not. On the filtered axis the design is competitive at high
$\sigma_{\mathrm{loc}}$ with Filtered-DiskANN ($86.56$ vs.\ $85.9$--$87.8$) while additionally supporting streaming
and multi-tenancy, and the low-$\sigma_{\mathrm{loc}}$ collapse is a shared graph-substrate property.

\section{Supplementary figures}
\label{sec:supp-figs}

\begin{figure}[htbp]\centering\includegraphics[width=0.7\columnwidth]{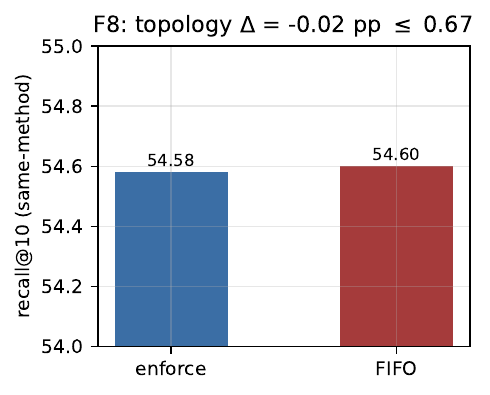}
\caption{Terminal recall under enforce- and FIFO-ordered merges on the identical $18$K terminal; $|\Delta|=0.02$ pp inside the $\le0.67$ pp band.}\label{fig:S1}\end{figure}

\begin{figure}[htbp]\centering\includegraphics[width=0.7\columnwidth]{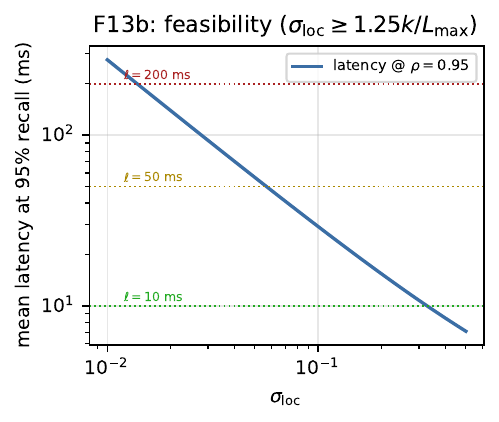}
\caption{Feasibility boundary: the mean latency required to reach $\rho{=}0.95$ (at
$L_{95}=1.25k/\sigma_{\mathrm{loc}}$) vs.\ local density, with candidate SLO lines $\ell\in\{10,50,200\}$ ms; a
tenant is feasible where its curve sits below its $\ell$. Low-$\sigma_{\mathrm{loc}}$ tenants (many-tenant,
diffuse) are infeasible at an interactive SLO; this is the priced infeasibility boundary. The linear-in-$L$ cost
model underlying the required-latency axis is Fig.~2 of the main text.}\label{fig:S2}\end{figure}

\begin{figure}[htbp]\centering\includegraphics[width=0.75\columnwidth]{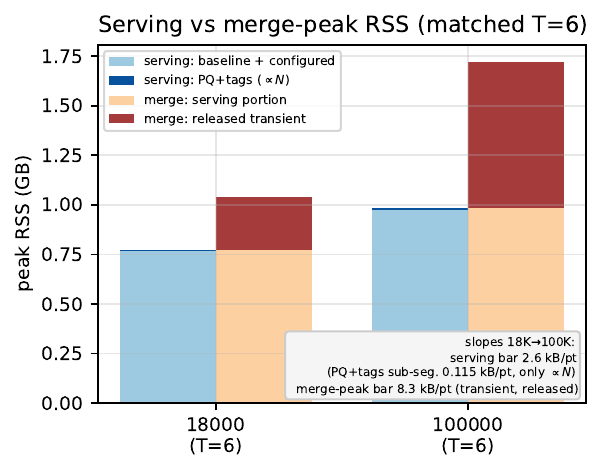}
\caption{RSS decomposition on the matched $T{=}6$ pair ($18$K, $100$K), on the conventions of
Table~IV. The \emph{serving} bar is whole-process first stable RSS, rising at $2.6$ kB/point across
the pair; its only $\propto N$ component is the PQ-codes-and-tags residency (graph on disk), the sub-segment
rising at $0.115$ kB/point that is the bounded quantity restoring the serving-memory claim, the remainder being a
fixed baseline plus configured, high-water-bounded allocations. The \emph{merge-peak} bar adds a released
transient (the StreamingMerger's in-RAM $O(N)$ rewrite plus build-thread arenas), giving an $8.3$ kB/point peak
that is reclaimed after the merge and is reducible by thread and batch configuration.}\label{fig:S3}\end{figure}